%% file: main.tex
\documentclass[a4paper,11pt]{article}
\usepackage{jheppub} 
\usepackage{lineno, verbatim}
\usepackage{physics}
\usepackage{bm}
\usepackage{subfig}
\usepackage{cancel}
\usepackage{comment}
\usepackage{mathtools}
\usepackage{empheq}
\usepackage{tikzit}
\input{Tikz/mlp_styles.tikzstyles}

\usepackage[normalem]{ulem}
\newcommand{\eec}{\text{EEEC}}

\newcommand{\cL}{\mathcal{L}}
\newcommand{\mtmc}{m_t^{\text{MC}}}

\newcommand{\bphi}{{\bm \phi}}

\title{\boldmath 
Observable Optimization for Precision Theory:
Machine Learning Energy Correlators}

\author[a]{Arindam Bhattacharya,}
\author[b,c]{Katherine Fraser,}
\author[a,d]{and Matthew D.~Schwartz} 

\affiliation[a]{Department of Physics, Harvard University, Cambridge, MA 02138, USA}
\affiliation[b]{Berkeley Center 
for Theoretical Physics, University of California, 
Berkeley, CA 94720, USA}
\affiliation[c]{Theoretical Physics Group, Lawrence Berkeley National Laboratory,  
Berkeley, CA 94720, USA}
\affiliation[d]{NSF Institute for Artificial Intelligence and Fundamental Interactions}

\emailAdd{arindamb@g.harvard.edu}
\emailAdd{kfraser@berkeley.edu}
\emailAdd{schwartz@g.harvard.edu}

\abstract{
 The practice of collider physics typically involves 
 the marginalization of multi-dimensional collider data to  uni-dimensional observables relevant for some physics task. In many cases, such as classification or anomaly detection, the observable can be arbitrarily complicated, such as the output of a neural network. However, for precision measurements, the observable must correspond to something computable systematically beyond the level of current simulation tools. 
 In this work, we demonstrate that precision-theory-compatible observable space exploration can be systematized by using neural simulation-based inference techniques from machine learning. We illustrate this approach by exploring the space of marginalizations of the energy 3-point correlator to optimize sensitivity to the the top quark mass.
We first learn the energy-weighted probability density from simulation, then search in the space of marginalizations for an optimal triangle shape. 
Although simulations and machine learning are used in the process of observable optimization, the output is an observable definition which can be then computed to high precision and compared directly to data without any memory of the computations which produced it.
We find that the optimal marginalization is isosceles triangles on the sphere with a side ratio approximately $1:1:\sqrt{2}$ (i.e. right triangles) within the set of marginalizations we consider.
}

\begin{document}
\maketitle
\flushbottom

\input{intro}

\section{Learning the Energy-Energy Correlator Distribution}\label{sec:eec_density}

In this section, we will describe our approaches using ML to learn the EEEC distribution. We study two different architectures: one a simple dense neural network (DNN), and the other a normalizing flow (NF). Before discussing the details of each network, we start by precisely defining the $\eec$.  

The three point energy correlator  from an $e^+e^-$ collision is defined as 
\begin{align}
    \eec (\zeta_1, \zeta_2, \zeta_3) &= \frac{1}{\sigma}\frac{d^3\sigma}{d\zeta_1 d\zeta_2 d\zeta_3}\ ,\nonumber\\ &= \frac{1}{\sigma}\frac{1}{2Q^2}\sum_{(i,j,k)}\int d\Pi_{n}~\left\vert\mathcal{M}(e^+e^- \rightarrow B)\right\vert^2 \frac{E_{i}E_{j}E_{k}}{Q^3} \; \delta_{ijk}(\text{shape}), \label{eq:eec_def}
\end{align}
where $(i,j,k)$ represents a triplet of particles in the final state $B$, $E$ denote the energies of those final state particles, $Q$ denotes the center of mass energy of the underlying collision process, $\sigma$ denotes the total integrated cross section of the process, and 
\begin{equation}
    \delta_{ijk}(\text{shape}) = ~\delta\left(\zeta_1 - \frac{1-\cos{\theta_{jk}}}{2}\right)~\delta\left(\zeta_2 - \frac{1-\cos{\theta_{ki}}}{2}\right)~\delta\left(\zeta_3 - \frac{1-\cos{\theta_{ij}}}{2}\right)\ .
\label{eqn:delta_shape}
\end{equation}
determines which triplets are included in the sum based on their relative angles. Here, $\zeta$ is a function of the pairwise angular separation $\theta$ of the particles. The EEEC as defined is a IRC safe quantity.\footnote{We restrict to the IRC safe EEECs, where the exponent of the energy weight is 1. More general EEECs can be defined by modifying that exponent, but such quantities are collinear unsafe.} Note that while each triplet is selected from a single jet, the sum is over both all triplets in a given jet and over all jets. Therefore,  EEECs are functions of an ensemble of events, and cannot be defined on an event-by-event basis.

\subsection{Simulation Details}
Our study is performed on a dataset of hadronically-decaying top jets arising in $t\bar{t}$ events from $e^+e^-$ collisions at centre-of-mass energy of $Q = 2$ TeV that is generated with \textsc{Pythia} 8.309 \cite{Bierlich:2022pfr}. 
The events were clustered
using anti-$k_t$ jets of radius $R=1.2$ as implemented in \textsc{Fastjet} 3.4.1 \cite{Cacciari:2005hq}. In order to reduce the large multiplicity in the resultant jets (the typical multiplicity of the anti-$k_t$ jets is around 150), the consituents were then reclustered via the Cambridge-Aachen algorithm~\cite{Dokshitzer:1997in,Wobisch:1998wt}  with radius $R^\prime =0.1$. This reduces the multiplicity to around 25.
In addition, we placed a lower cutoff on the product of the energies of each triplet, selecting only those with $\frac{E_i E_j E_k}{Q^3}\geq 10^{-6}$. This cut significantly speeds processing and has little effect on the resulting EEECs. Then for each jet, we look at the roughly $\binom{25}{3}  \sim 5000$ tuples and measure the angles and energies. This produces a distribution of points
\begin{equation}
\vec{x} = \left(\zeta_{1},\zeta_{2},\zeta_{3},\frac{E_1 E_2 E_3}{Q^3}, m_t\right),
\label{tuples}
\end{equation} 
in a 5 dimensional space. 
Here $\zeta_j$ are functions of the relative angles between particles, as in \eqref{eqn:delta_shape},
which are sorted so that $\zeta_1\le \zeta_2 \le \zeta_3$.
$m_t$ is Monte Carlo mass $\mtmc$, a parameter of the simulation. We generate samples with $m_t$ between $170$ and $180$ GeV, with a spacing of $0.1$ GeV. At each $\mtmc$ we generate 1M jets, and use a subset of them for training depending on the architecture. 

\subsection{Learning a Probability Distribution}
Having simulated the events necessary to compute the EEECs, we next wish to learn the distribution of the simulated samples. 
We would like to train a neural network probability density $p_\phi(\vec{x})$ which models the distribution of points $\vec{x}$ generated in  the simulation.

The task of approximating a probability density using a large number of samples drawn from a distribution is a well-known statistical problem. 
A standard approach is to minimize the Kullback-Leibler (KL)  divergence~\cite{kl_div} between the underlying data density $p_{\text{Data}}(\vec{x})$ and a normalizable density function $p_{\bm{\phi}}(\vec{x})$ parametrized by the neural network weights $\bm{\phi}$. 
Namely, the loss function $\mathcal{L}(\bm{\phi})$ is given by
\begin{align}
    \mathcal{L}_{\rm KL}(\bm{\phi}) &= D_{\rm KL}(p_{\text{Data}}(\vec{x})||p_{\bm{\phi}}(\vec{x}))\geq 0\ , \nonumber\\
    &=-\int d^{n}\vec{x}~p_{\text{Data}}(\vec{x})\ln(p_{\bm{\phi}}(\vec{x})) + \rm{constant}\ ,\nonumber\\
    &=-\mathbb{E}_{p_{\rm Data}(\vec{x})}\left[\ln {p_{\bm \phi}(\vec{x})}\right] + \rm{constant}\  \ , \label{eq:KL_Loss}
\end{align}
where the constant term 
does not depend on the weights $\bm \phi$. Since the KL divergence is minimized only when $p_{\text{Data}}=p_{\bm{\phi}}$, one is guaranteed that $p_{\bm{\phi}}$ will converge to $p_{\text{Data}}$ in the limit of infinite statistics and perfect training. 
In practice one has to estimate the loss not as a continuous integral but as a discrete average over samples $\{\vec{x}_{i}\}_{i=1}^{N}$ in training data. Namely,
\begin{align}
    \mathcal{L}_{\rm NLL}(\bm{\phi}) &\approx -\frac{1}{N}\sum_{i=1}^{N} \ln{p_{\bm \phi}(\vec{x}_{i})}\ , \label{eq:nll_loss}
\end{align}
which reduces the KL loss to the 
negative log likelihood (NLL) loss, which has been used for various density estimation tasks in particle physics. 
Minimizing the loss as written in Eq.~\eqref{eq:nll_loss} is an ill-posed problem, since one can arbitrarily decrease the loss by continuously increasing the density every time a new data point is seen. 
Thus, the loss function in Eq.~\eqref{eq:nll_loss} needs to be regularized by restricting the function space of parametrized $p_{\bm \phi}(\vec{x})$. There are several different ways of regularizing $p_{\bm \phi}(\vec{x})$ which enforce its finite normalization; we regularize each of our architectures in distinct ways as described below.

\subsection{DNN Density Approach}\label{sec:dnn_density}
 Our first density estimation approach is a dense neural network (DNN) which outputs $p_{\bm \phi}(\vec{x})$. In this case, we regularize the NLL loss by adding an integral over the probability density to explicitly enforce the normalization constraint. Specifically, we modify the loss to be
\begin{align}
    \mathcal{L}_{\rm NLL, \lambda}(\bm{\phi}) &=-\mathbb{E}_{p_{\rm Data}(\vec{x})}\left[\ln {p_{\bm \phi}(\vec{x})}\right] + \lambda\ \bigg\vert\ln \int d^m x~ p_{\bm \phi}(\vec{x})\bigg\vert\\
    &\approx -\frac{1}{N}\sum_{i=1}^{N} \ln{p_{\bm \phi}(\vec{x}_i)}\ + \lambda\bigg\vert \ln(\frac{1}{M}\sum_{j=1}^{M} p_{\bm \phi}(\vec{x}_{j}))\bigg\vert\ .
    \label{eq:nll_reg}
\end{align}
It is the second term which imposes the normalization constraint on the learnt function, and focuses the optimization over $L^{1}$ functions. The discrete sum over $N$ and $M$ occurs over the batch used during training. 

Computing this integral explicitly can be computationally intractable for high dimensional distributions, but is feasible in our case of tuples $(\vec{\zeta}, \tilde{E},m_{t})$ in $m=5$ dimensions (see Eq.~\eqref{tuples}).  
Here, $\tilde{E} = E_1 E_2 E_3/Q^3$ is the normalized product of energies of particles within a tuple and $\vec{\zeta}=(\zeta_1,\zeta_2, \zeta_3)$ are
functions of the angles on the sphere where the EEEC is measured, as defined in Eq.~\eqref{eqn:delta_shape}. We compute the integral in
Eq.~\eqref{eq:nll_reg} directly using Monte Carlo integration with the library \verb|torchquad|~\cite{Gómez2021} 
and set $\lambda$ to $100$.

Once the DNN converges to a $p_{\bphi}(\vec{\zeta},\tilde{E},m_{t})$, one 
can obtain the full four dimensional EEEC by integrating over $\tilde{E}$, i.e.  
\begin{align}
    \eec_{\bm \phi}(\vec{\zeta},m_{t}) = \int d\tilde{E}~\tilde{E}~p_{\bm \phi}(\vec{\zeta},\tilde{E},m_{t}) \label{eq:eec_int}
\end{align}
Once $\eec_{\bm \phi}(\vec{\zeta},m_{t})$ is known, marginalized lower dimensional EEECs from which we can extract the top-quark mass can be computed by additionally integrating $\eec_{\bm \phi}(\vec{\zeta},m_{t})$ over selected regions in $\vec{\zeta}$ space.

Unfortunately, while it is possible to learn $p(\vec{\zeta}, \tilde{E},m_{t})$, the resulting
distribution learned by the DNN is most accurate in the regions that have the highest probability density, which is where there are multiple soft (low-energy) particles. However, these carry little weight in Eq.~\eqref{eq:eec_int}.
Conversely, the learnt density $p_{\bm \phi}(\vec{\zeta},\tilde{E},m_{t})$ and therefore $\eec_{\bm \phi}(\vec{\zeta},m_{t})$ generally struggles to capture the high energy region of the distribution that is needed for $m_t$ discrimination.

It becomes apparent that the DNN learning the probability distribution is not ideal. Instead,
one needs to focus the network's attention to the high-energy region of interest in the distribution.
To do so, we observe from Eq.~\eqref{eq:eec_int} that 
the EEEC is simply a positively re-weighted four-dimensional distribution.
Although it is not a probability density, one can learn it with the same techniques as are used to learn a probability density:
compute the KL divergence between the EEEC as given in the data and a neural net parametrization $\eec_{\bm \phi}$, namely
\begin{align}
    \mathcal{L}_{\text{EEEC}}(\bm \phi) &= D_{\text{KL}}(\eec_{\text{Data}}(\vec{\zeta},m_{t})||\eec_{\bm \phi}(\vec{\zeta},m_{t}))\\
    &=-\mathbb{E}_{\eec_{\text{Data}}}[\ln{\eec_{\bm \phi}(\vec{\zeta},m_{t})}] +\text{constant}\\
    &= -\int d^{3}\vec{\zeta}~dm_{t}~\eec_{\rm Data}(\vec{\zeta},m_{t})~\ln{\eec_{\bm \phi}(\vec{\zeta},m_{t})} +\text{constant}\\
    &=-\int d^{3}\vec{\zeta}~dm_{t}\left(\int d\tilde{E}~\tilde{E}~p_{\text{Data}}(\vec{\zeta},\tilde{E},m_{t}) \right)~\ln{\eec_{\bm \phi}(\vec{\zeta},m_{t})} +\text{constant}
\end{align}
The discretized version of the loss function that we use in practice is
\begin{align}
    \mathcal{L}_{\text{EEEC}, \lambda}(\bm \phi) &= -\mathbb{E}_{\eec_{\text{Data}}}[\ln{\eec_{\bm \phi}(\vec{\zeta},m_{t})}]+ \lambda~\bigg\vert\ln(\int d^{3}\vec{\zeta}~dm_t~
    \eec_{\bm \phi}(\vec{\zeta},m_t))\bigg\vert\\\
    &\approx - \frac{1}{N}\sum_{i=1}^{N} \tilde{E}_{i} \ln{\eec_{\bm \phi}(\vec{\zeta}_{i},m_{t, i})} + \lambda~\bigg\vert\ln(\frac{1}{M}\sum_{j=1}^{M}
    \eec_{\bm \phi}(\vec{\zeta}_{j},m_{t, j}))\bigg\vert\ ,\label{eq:eec_loss}
\end{align}
 Convergence of the loss function (assuming infinite statistics and perfect training) in Eq.~\eqref{eq:eec_loss} is guaranteed as it is based on the KL divergence between $\eec_{\rm Data}$ and $\eec_{\phi}$. Additionally, $\mathcal{L}_{\eec, \lambda}(\bm \phi)$ is physically intuitive, since it not only prioritizes regions where samples are concentrated, but also re-weights them by their energy like the original observable. As with learning the ordinary probability distribution, the normalization term in Eq.~\eqref{eq:eec_loss} is needed to regularize the loss function and make the optimization problem well-posed. We remind the reader that the discrete sum over $N$ and $M$ occurs over the batch used for training the network 

\subsubsection{Training Details}
Effectively training the DNN requires preprocessing the simulated EEEC tuples. Since neural networks are sensitive to the order of magnitude of their inputs and training data, we employed bijective normalizations to the tuples to get $\mathcal{O}(1)$ numbers. Specifically, we first scale the sides $\vec{\zeta}$ and energy products $\tilde{E}$ by a log transformation, and then scale the transformed inputs and $m_t$ to the interval $[0,1]$. The composed transformation is
\begin{align}
    \hat{\zeta}_i &= \frac{\log_{10}(\zeta_{i}/\zeta_{i, \min})}{\log_{10}(\zeta_{i, \max}/\zeta_{i, \min})}\ ,  & \hat{m}_{t} &=\frac{m_{t}-m_{t,\min}}{m_{t,\max}-m_{t,\min}}\ ,   \label{eq:norm_transform}
\end{align}
where $i\in\{1,2,3\}$. In addition, another transformation where we excluded the tail of $\vec{\zeta}$ was performed, amounting to excising the interval in $\hat{\zeta}_i \in [0,1]$ where the probability fell below $\sim 1\%$.  Post excision, the $\hat{\zeta}$ variables were again rescaled to $[0,1]$ using a linear transformation. The exclusion of the tails is beneficial because it improves training without affecting the region of the EEEC that is sensitive to the top mass (which arises from intermediate $\vec{\zeta}$). 

Following preprocessing, we train a 7 layer deep multi-layer-perceptron (MLP) network with 256 nodes each and ReLU activation, with the final layer outputting $\ln({\eec}_{\bm \phi})$. Learning the logarithm of the EEEC is more numerically stable and one can easily exponentiate the forward pass of the trained MLP to get $\eec_{\bm \phi}$. The training set consisted of triplets from 10000  jets at each mass, using a 90:10 train-validation split. Early stopping and reduce learning rate on plateau were employed to prevent overfitting, with the validation loss as the metric. Other training details are given in Table~\ref{tab:exp_details}. In order to get stable convergence, we also implemented stochastic weight averaging~\cite{Izmailov:2018} after 50 epochs, with cosine annealing of the learning rate over 20 epochs, and the employed learning rate for weight averaging was $8\times 10^{-7}$. 

\begin{table}[!tbp]
    \centering
    \begin{tabular}{|c|c|}
        \hline
        Learning Rate & $8\times 10^{-5}$\\
        \hline
        Training Batch Size & $4096$ \\
        \hline
        Optimizer & AdamW\\
        \hline
        Weight Decay & $10^{-8}$\\
        \hline
        Number of Epochs & $100$\\
        \hline
        Early Stop Patience & $20$\\
        \hline
        Learning Rate Drop Factor & $0.5$\\
        \hline
        Learning Rate Drop Epochs & $10$\\
        \hline
    \end{tabular}
    \caption{Architecture and Training Details for the MLP used to learn the EEEC density from simulated \textsc{Pythia} data.}
    \label{tab:exp_details}
\end{table}

\subsubsection{Testing Marginals}
\begin{figure}[!htbp]
    \centering
    \includegraphics[scale=0.7]{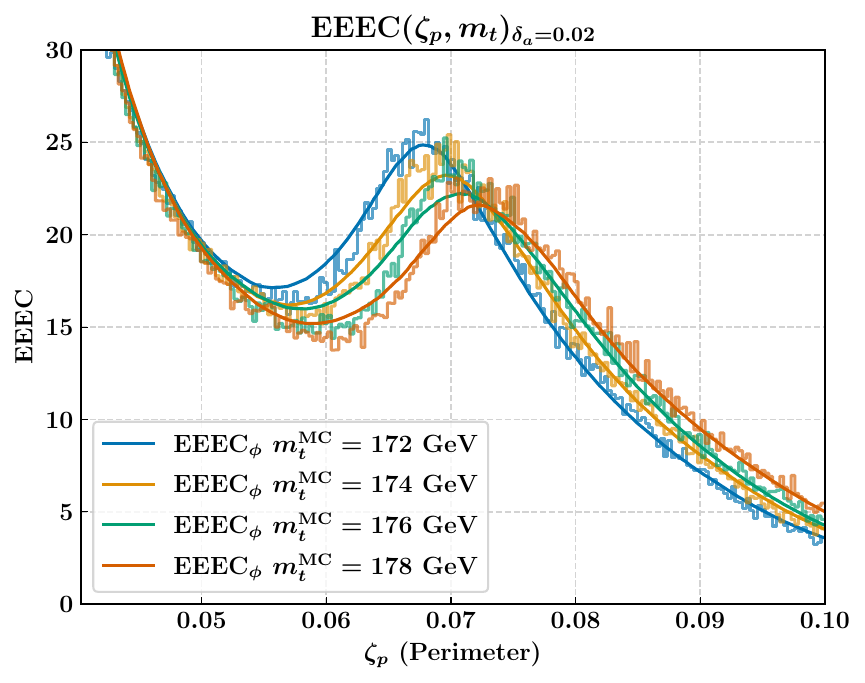}
    \caption{Comparison of the EEEC$_{\bm \phi}$ distribution learned by the DNN to that computed directly from \textsc{Pythia} data as a function of the perimeter of the triangle $\zeta_p = \zeta_1+\zeta_2+\zeta_3$ for `equilateral' triangles with the asymmetry parameter $\delta_{a} \leq 0.02$. The EEEC is normalized to integrate to 1 in the range shown.}
    \label{fig:equi_tri_mt_var}
\end{figure}

\begin{figure}[!htbp]
    \centering
    \includegraphics[width=\columnwidth]{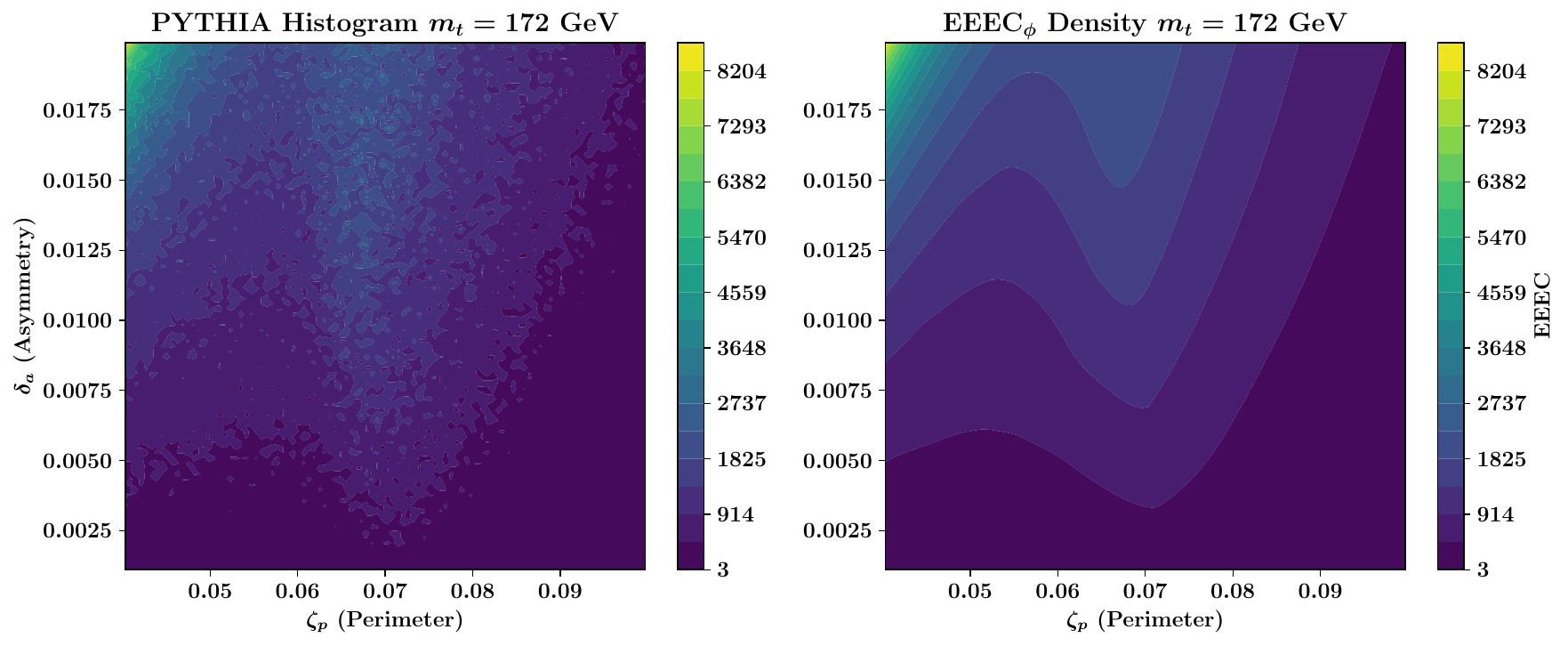}
    \caption{
    Comparison of a representative two dimensional differential EEEC distribution as a function of  $\zeta_p$ (triangle perimeter) and $\delta_{a}$ (asymmetry) between the learned  EEEC$_{\bm \phi}$ and \textsc{Pythia} histograms. Histograms are normalized to integrate to 1 over the plotted region. We find good agreement.}
    \label{fig:perim_asym_2d}
\end{figure}

To test that the learned density emulates the underlying EEEC distribution, we compare its various marginals to the EEECs computed directly from \textsc{Pythia}. For the data histograms, a larger dataset (1M) of events needs to be used in order to get a smooth distribution, since the histogram fit does not have data from other masses, unlike the NN. By marginalizing, we are able to design tests which focus on particular regions of parameter space, including those which are sensitive to the top-quark mass. Here, we work with two different parameterizations of the marginalization. First, we work with a parameterization in terms of perimeter ($\delta_p = \zeta_1+\zeta_2+\zeta_3$) and asymmetry ($\delta_{a}=\zeta_3-\zeta_1$). We consider this parameterization in order to compare to previous work \cite{Holguin:2022epo}. We verify that the network learns both equilaterial triangles, where the EEEC is a function of perimeter and $\delta_a \leq 0.02$, and the double differential distribution of both perimeter and asymmetry. These are shown in Figs.~\ref{fig:equi_tri_mt_var} and~\ref{fig:perim_asym_2d}, respectively. We find good agreement between data and the learned network for both.

\begin{figure}[!tbp] 
\centering 
\subfloat[]{ \includegraphics[width=0.48\columnwidth]{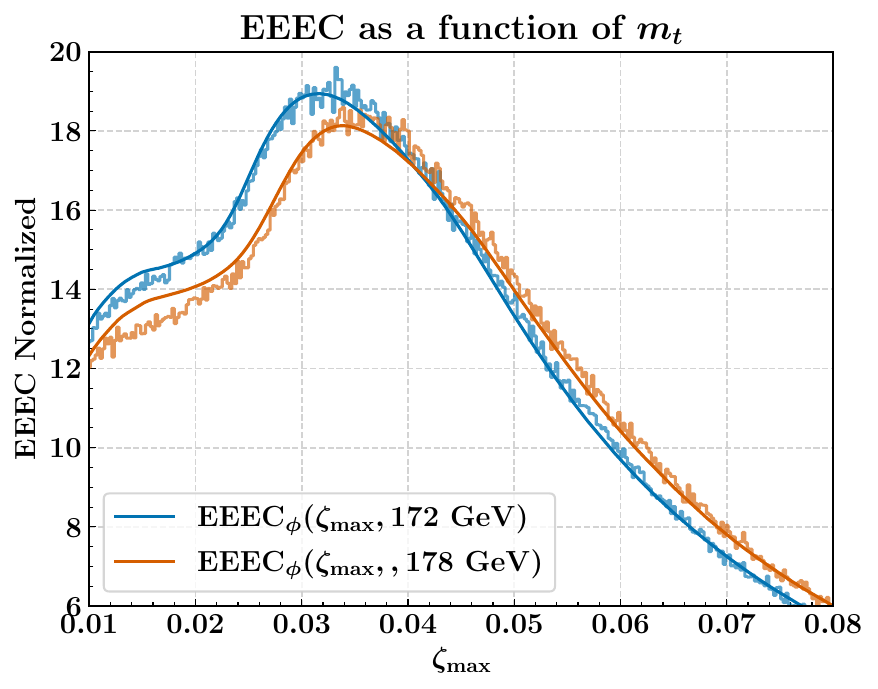} 
\label{fig:subshape1} 
} 
\subfloat[]{ \includegraphics[width=0.48\columnwidth]{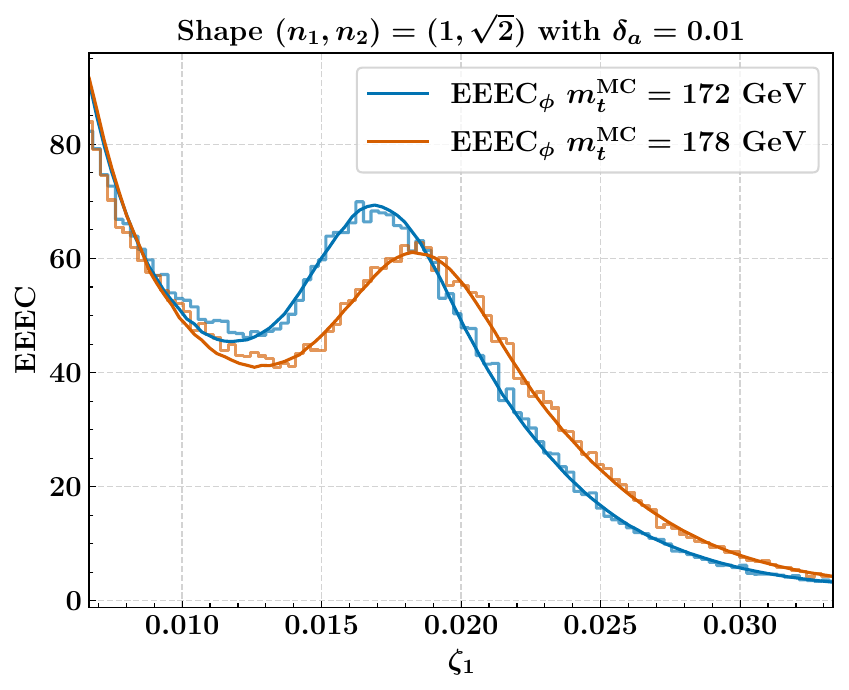}
\label{fig:subshape2}
} 
\caption{Comparing observables computed directly from \textsc{Pythia} data to marginals from the learnt EEEC. On the left, we plot the projected EEEC as a function of $\zeta_\text{max} = \zeta_3$, and on the right we plot the distribution of right-angled isosceles triangles. The learnt EEEC density model indeed produces an analytic surrogate that captures features of the underlying simulated data.} 
\label{fig:shape_compare} 
\end{figure}

The second marginalization we evaluate is the EEEC as a function of a single side. In the left panel of Fig.~\ref{fig:shape_compare}, we show the marginalization in the large $\delta_a$ limit down to the largest side $\zeta_{\text{max}}=\zeta_3$, which includes all events (as in \cite{Chen:2020vvp}).
In the right panel of Fig.~\ref{fig:shape_compare}, we plot a single example of the parameterization
\begin{multline}
        \eec(\zeta_1, n_1,n_2, m_t)_{\delta_{a}} = \int_0^{1} d\zeta_2~d\zeta_3~ \eec(\zeta_1,\zeta_2,\zeta_3, m_t)\\
        \times~\theta(\zeta_2-n_1\zeta_1)~\theta(\delta_a + n_1\zeta_1 -\zeta_2)~\theta(\zeta_3-n_2\zeta_1)~\theta(\delta_a + n_2\zeta_1 -\zeta_3),
        \label{eqn:shape}
    \end{multline}
though we also check other values of $n_1, n_2, \delta_a$ and also find good agreement in those cases. This corresponds to triangles which are roughly of the shape $\zeta_1 \times (1, n_1, n_2)$, up to a smearing window $\delta_a$. It is this parameterization in Eq.~\eqref{eqn:shape} that we will continue to use later to search the parameter space of different possible shapes.

\subsection{Normalizing Flows}\label{sec:nflow}

While the DNN architecture of Sec.~\ref{sec:dnn_density} is quite simple, the explicit calculation of the integral of the network output rapidly becomes computationally intractable as the dimension $m$ of the network increases. For example, with 
images, computing an integral over the density is prohibitive because $m$ scales as the number of pixels. With such data in mind, more sophisticated methods such as normalizing flows have been developed which do not rely on explicit computation of the integral. Such flows constrain the learnt density by imposing $p_\bphi(\vec{x})$ to be the image under a bijective map of a simple, normalizable base distribution such as a Gaussian. Like the dense network, flows use the KL divergence between $p_{\bm \phi}(\vec{x})$ and the data distribution as the loss function. Because the map is bijective, flows also allow sampling from the learned distribution, though we do not need sampling for our study.

Explicitly, suppose we start with a base distribution $p_0$ of samples $z_0$. Then a flow is obtained by successively apply bijective transformations $f_1$, $f_2$, \dots $f_{K-1}$ to get the final distribution $p_K$ of samples $z_K$, with $K$ a free parameter describing the number of transformations which is tuned for a specific application. In our case, $z_K$ consists of the five inputs we would like to learn the density of: the three $\zeta$'s, the product of energies $\tilde{E} = E_1 E_2 E_3$, and the top-quark mass. From this, density is simply evaluated with the chain rule to be
\begin{equation}
    \log(p_K(z_K)) = \log(p_0(f^{-1}(z_K))) - \log(\det \Big|\frac{df^{-1}(x)}{dx}\Big|)
\end{equation}
where we have defined $f \equiv f_{K-1} \circ f_{K-2} \dots \circ f_1$.
Because computing the density (and the loss) rely on the Jacobian, flow architectures are typically chosen so that Jacobians are easy to compute. A common choice that we employ is using an autoregressive network \cite{Papamakarios:2017tec,  Germain:2015yft}, where each layer depends on one additional input dimension, making the final transformation matrix triangular and therefore easy to compute the determinant. The network learns the parameters of a rational quadratic spline which is an invertible but expressive transformation \cite{nflows}. There are also many other existing flow architectures which we do not study, including inverse autoregressive flows \cite{Kingma:2016wtg} and continuous normalizing flows \cite{Chen:2018, Grathwohl:2018vdm}.

Similiar to the DNN case, once the probability distribution $p_{\bphi}(\vec{\zeta},\tilde{E},m_{t})$ has been obtained, the EEEC is obtained by integrating $\tilde{E} p_{\bphi}(\vec{\zeta},\tilde{E},m_{t})$ over $\tilde{E}$ (see Eq.~\ref{eq:eec_int}). Also as before, lower dimensional differential EEECs can then be obtained by integrating over selected regions of $\vec{\zeta}$ space, as in Eq.~\ref{eqn:shape}. Unlike in the DNN case, the flow is also able to accurately learn the parts of the probability distribution which are sensitive to the top-quark mass without needing to energy weight the loss function. This is one advantage of the more complicated, specialized flow architecture over the much simpler DNN. 

\subsubsection{Training Details}

Effectively training the flow also requires order one inputs. For the flow, we use the same mapping on the top-quark mass as in Eq.~\ref{eq:norm_transform}, and use 
\begin{align}
    \hat{\zeta}_i &= \ln(\frac{\zeta_{i}/(1.1\zeta_{\mathrm{max}})}{1 - \zeta_{i}/(1.1\zeta_{\mathrm{max}})})\ ,  & \hat{E} &=\ln(\frac{\tilde{E}/(1.1 \tilde{E}_{\mathrm{max}})}{1 - \tilde{E}/(1.1 \tilde{E}_{\mathrm{max}})})\ , 
    \label{eq:norm_flow}
\end{align}
for the product of energies $\tilde{E}$ and the angles $\zeta_i$. Note that $\zeta_{max}$ is the maximum of all the $\zeta_i$ for $i \in {1, 2, 3}$, and is the same for all three maps. 

Our flow is a five dimensional, ten block deep flow implemented using \verb|nflows|\cite{nflows} and modeled on the architecture from \cite{Krause:2021ilc, Shih2022}. Each block consists of a rational quadratic spline (RQS) implemented using  Masked Piecewise Rational Quadratic Autoregressive Transforms, a batch normalization layer, and a random permutation. For the RQS layers, we also use: ReLU activations, 40 bins, 200 hidden features, 2 context features, a tail bound of 14, min bin widths of $10^{-6}$, linear tails, and turn off residual blocks.\footnote{We also tested flows with residual blocks as in \cite{Behrmann_invert_residual_net}, but found they both slowed training and decreased performance.} The base distribution is taken to be Gaussian in the four dimensions corresponding to energy and $\zeta$, and uniform between $[-1,2]$ in the dimension corresponding to $m_t$ to avoid edge effects. The network was trained for 313 epochs (9 days on our GPU), where each epoch consists of reading 500,000 tuples randomly sampled from each mass (with tuples computed from the full 1M events), and each training batch consists of 500,000 tuples randomly mixed between masses.
The flow is trained using the Adam optimizer with initial learning rate $10^{-4}$, with the learning rate reduced by a factor of 2 after 20 epochs without improvement, and early stopping after 50 epochs without improvement. We also tested energy weighting the flow loss, but it did not noticeably improve performance.

\subsubsection{Testing Marginals}

\begin{figure}[!t] 
\centering 
\subfloat[]{\includegraphics[width=0.63\textwidth, height=5.6cm, keepaspectratio]{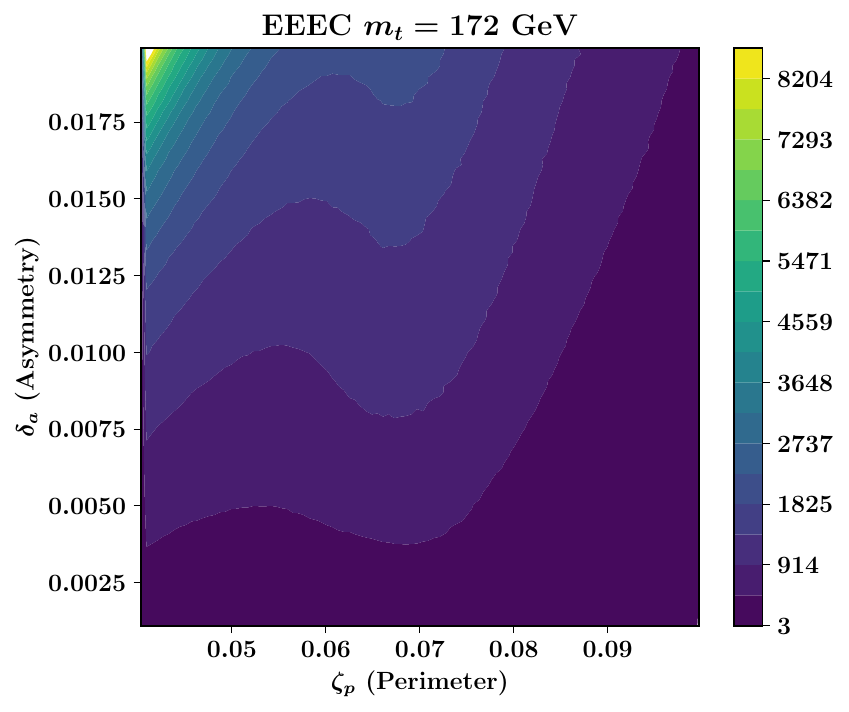} 
\label{fig:shape1} 
} 
\hfill
\subfloat[]{\includegraphics[width=0.47\textwidth, height=6cm, keepaspectratio]{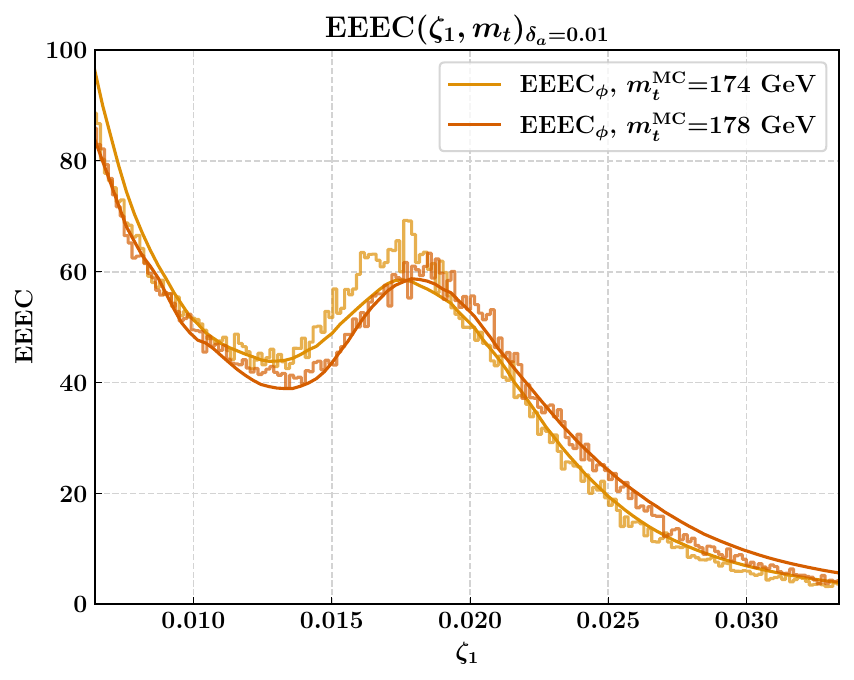}
\label{fig:shape3}
}
\caption{Shape distributions learned by the normalizing flow. Left: Two dimensional marginal to asymmetry and perimeter $\zeta_p = \zeta_1 + \zeta_2 + \zeta_3$. Right: Shape from Eq.~\ref{eqn:shape} with $(n_1, n_2, \delta_a) = (1, \sqrt{2}, 0.01)$. Both offer reasonable agreement with data, but the agreement with data is not as good as for the DNN (see Figs.~\ref{fig:perim_asym_2d} and~\ref{fig:subshape2}).} 
\label{fig:flow_shapes} 
\end{figure}

Here we show the results of training the flow. Specifically, we show two different marginalizations in Fig~\ref{fig:flow_shapes}: the residuals between data and the flow as a function of perimeter $\zeta_p$ and asymmetry $\delta_a$ and the shape defined in Eq.~\ref{eqn:shape} with $n_1 = 1$, $n_2 = \sqrt{2}$, $\delta = 0.01$. These are the same shapes as shown for the DNN case in Figs.~\ref{fig:perim_asym_2d} and~\ref{fig:subshape2}.

As can be seen from Fig~\ref{fig:flow_shapes}, both shapes agree reasonably well with the data. However, neither shape agrees with the data quite as well as those learned by the DNN. It is possible this is due to the details of our architecture, and that additional engineering will improve the shapes learned by the flow. This additional engineering might be advantageous if there are cases where we wish to learn the high energy part of the probability distribution, instead of its energy weighted version. However, we do want the energy weighted version, so we choose to focus on the DNN for the rest of the paper, both because of its better agreement with data and because the flow's complexity makes it slower to evaluate.

\section{Top Mass from EEECs: An Application of Neural Ratio Estimation}\label{sec:fit}
Having learnt the fully differential EEEC distribution, we now seek to find a marginalization that optimizes the sensitivity to the top mass. To begin,
we need a method to regress the top mass from a marginal (or equivalently, shape) of the analytic surrogate $\eec_{\bm \phi}(\vec{\zeta}, m_t)$ which also provides an uncertainty or confidence range on this mass estimate.

\emph{Neural ratio estimation} (NRE) ~\cite{Hermans:2019,Cranmer:2015bka,Cole:2021gwr} is one such method.
Given a prior on the top mass, NRE allows one to extract a posterior for the top mass for a given EEEC marginal or shape. The posterior provides both an estimate for  the top mass in the form of its mode a.k.a the maximum a-posteriori (MAP), and an uncertainty estimate in the form of its width. 
Thus, with access to posteriors for multiple shapes, we can statistically assess the sensitivity of shapes or marginals to the underlying top mass, and rigorously compare whether one is better than the other for mass regression. In what follows, we focus on the shapes or marginals defined in equation~\ref{eqn:shape}, computed with the DNN since it is faster to evaluate than the flow. Recall that these probe triangles roughly of the form $\zeta_1 \times (1, n_1, n_2)$ with a smearing window of $\delta_a$.
We show how one can use NRE to compare different shapes, and optimize the shape with respect to $m_{t}$ regression in a region of $(n_1, n_2)$. The methods that follow could also be applied to more complicated shapes, higher dimensional marginals and other functions of the full $\eec$ distribution, but we stick to this parameterization for simplicity.

\subsection{Architecture and Training \label{sec:NRE_arch}} 

NRE aims to extract the posterior for the underlying top mass given a certain shape or marginalization of the EEEC. It does this while avoiding both the challenge of explicitly writing a tractable likelihood, and the computational difficulty of evaluating the posterior. NRE gets around these problems by using the ratio trick to directly learn $\frac{p(\vec{\zeta}_\text{shape},m_t)}{p(\vec{\zeta}_\text{shape}) p(m_t)}$ using a parametrized classifier. By definition, one has that 
\begin{align}
    \frac{p(\vec{\zeta}_\text{shape},m_t)}{p(\vec{\zeta}_\text{shape}) p(m_t)} = \frac{p(\vec{\zeta}_{\text{shape}}\vert m_t)}{p(\vec{\zeta}_\text{shape})} = \frac{p(m_t\vert\vec{\zeta}_{\text{shape}})}{p(m_t)}
\end{align}
where the last term is the ratio of the posterior to the prior. A classifier learns to discriminate between the joint distribution $p(\vec{\zeta}_\text{shape},m_t)$ and the product of the marginal distributions $p(\vec{\zeta}_\text{shape}) p(m_t)$ using the binary cross entropy loss function 
\begin{align}
    \cL_{\text{NRE}}(\bm{\phi}) = -\sum_{i=1}^{N}\left[p(\vec{\zeta}^{i}_\text{shape},m_t^{i}) \ln[\sigma(f_{\bm{\phi}})]+p(\vec{\zeta}^{i}_\text{shape})p(m_t^{i}) \ln[1-\sigma(f_{\bm{\phi}})]\right] 
    \label{eq:nre_loss}
\end{align}
where $\sigma$ is the sigmoid function and $f_{\bm{\phi}}$ denotes the neural net classifier with parameters $\bm{\phi}$. Here, one assigns a label $1$ to samples from the joint distribution $p(\vec{\zeta}_\text{shape},m_t)$ and label $0$ to samples $\vec{\zeta}_\text{shape}$ with arbitrary labels drawn from $p(m_t)$. 
The function that minimizes the loss in Eq.~\eqref{eq:nre_loss} satisfies 
\begin{align}
    f_{\bm{\phi}}(\vec{\zeta}_\text{shape},m_t) &= \ln(\frac{p(\vec{\zeta}_\text{shape},m_t)}{p(\vec{\zeta}_\text{shape}) p(m_t)}) = \ln(\frac{p(m_t\vert\vec{\zeta}_{\text{shape}})}{p(m_t)})
\end{align}
Therefore, a forward pass of the ideal classifier provides access to the posterior $p(m_t\vert \vec{\zeta}_{\text{Shape}})$ for a given prior $p(m_t)$. 

\begin{figure}[!t]
  \centering
  \scalebox{1.15}{\input{Tikz/mlp_example.tikz}}
    \caption{A schematic of the NRE architecture used to obtain the posterior $p(m_{t}\vert \vec{\zeta}_{\rm Shape}) = p(m_{t})\exp(f_{\bm \phi}(\vec{\zeta}_{\rm Shape}, m_{t}))$, including a summary network $s_{\bm \phi}$ for the shape $\vec{\zeta}_{\rm Shape}$ and a classifier network $c_\phi$. Gaussian noise $\epsilon$ at $5\%$ level was added to the density curves of the shapes to mimic histogram sampling noise, as described in Eqn.\eqref{eq:gauss_noise}.}
  \label{fig:MLP_cartoon}
\end{figure}

\begin{table}[!htbp]
    \centering
    \begin{tabular}{|c|c|c|}
        \hline
        Network & $s_{\bm{\phi}}$ & $c_{\bm{\phi}}$ \\
        \hline
        No. of Layers & 5 & 5\\
        \hline
        Features for Layers & [100,128,128,128,128] & [4,128,128,128,128]\\
        \hline
        Dropout & [0,0.01,0.02,0.04,0.08] & [0,0,0,0,0]\\
        \hline
        Activation & ELU, $\alpha=1$ & ELU $\alpha=1$ \\
        \hline 
    \end{tabular}
    \caption{Architecture details for the NRE network. Both networks were trained simultaneously using a learning rate of $10^{-3}$ and a batch-size of 2000.}
    \label{tab:arch_nre}
\end{table}

In practice, we parametrize the shape as a discrete set of 100 uniformly spaced points of the function $\eec(\zeta_1,n_1, n_2, m_t)_{\delta_a=0.01}$, with $3\zeta_1\in[0.02,0.1]$, normalized such that $\sum_{i}\eec(\zeta^{i}_1,n_1, n_2, m_t)_{\delta_a=0.01}=1$. In order to simplify the classification, we also use a summary network to condense the information in the shape to a single number, using an MLP $s_{\bm{\phi}}$~\cite{Cole:2021gwr}. Specifically, the inputs to the summary network are  
\begin{multline}
     \vec{\zeta}_\text{shape} = [\eec(\zeta^{i=0}_1,n_1, n_2, m_t)_{\delta_a=0.01}(1+\epsilon_{i=0}), \eec(\zeta^{i=1}_1,n_1, n_2, m_t)_{\delta_a=0.01}(1+\epsilon_{i=1}),\\ \cdots, \eec(\zeta^{i=100}_1,n_1, n_2, m_t)_{\delta_a=0.01}(1+\epsilon_{i=100})] \label{eq:gauss_noise}
\end{multline}
where the $\epsilon_{i}$ is Gaussian random noise at 5$\%$ level of the $i^{\text{th}}$ bin. Since the learnt $\eec_{\bm \phi}$ outputs the density, noise was added to mimic sampling error for resultant histograms. Thereafter, the classification uses another MLP network $c_{\bm{\phi}}$ with inputs $m_t$, $n_1$, $n_2$, $s_{\bm{\phi}}(\vec{\zeta}_{\text{shape}})$. A schematic visualization of this architecture is given in Figure~\ref{fig:MLP_cartoon}, and more details are presented in Table~\ref{tab:arch_nre}.

Our training data consisted of a training set of $9\times 10^{5}$ shapes and a validation set of $10^5$ shapes, with the top mass uniformly drawn from $[170,179.9]$ GeV. Shape parameters $n_1, n_2$ were drawn from an exponential distribution, assuming $n_2\geq n_1$. The top mass has again been rescaled to $[0,1]$ using Eq.~\eqref{eq:norm_transform}, and similarly the shape parameters are fed as input to the net after taking their natural logarithm. 
Both the summary network $s_\phi$ and the classification network $c_\phi$ are trained simulatenously for 600 epochs, with stochastic weight averaging being implemented after 480 epochs with cosine annealing of the learning rate over 20 epochs, and early stopping tracking the validation loss implemented in order to prevent overfitting.

\subsection{Comparing Shapes} 

Next we seek to compare different shapes using the NRE architecture described in Sec.~\ref{sec:NRE_arch}. In order to do this, we need a way to define a metric with which to quantify the error and check that it is reliable. Then we need to evaluate the trained network on many different shapes in order to minimize this error.

\begin{figure}[!t] 
\centering 
\subfloat[]{ \includegraphics[width=0.5\textwidth]{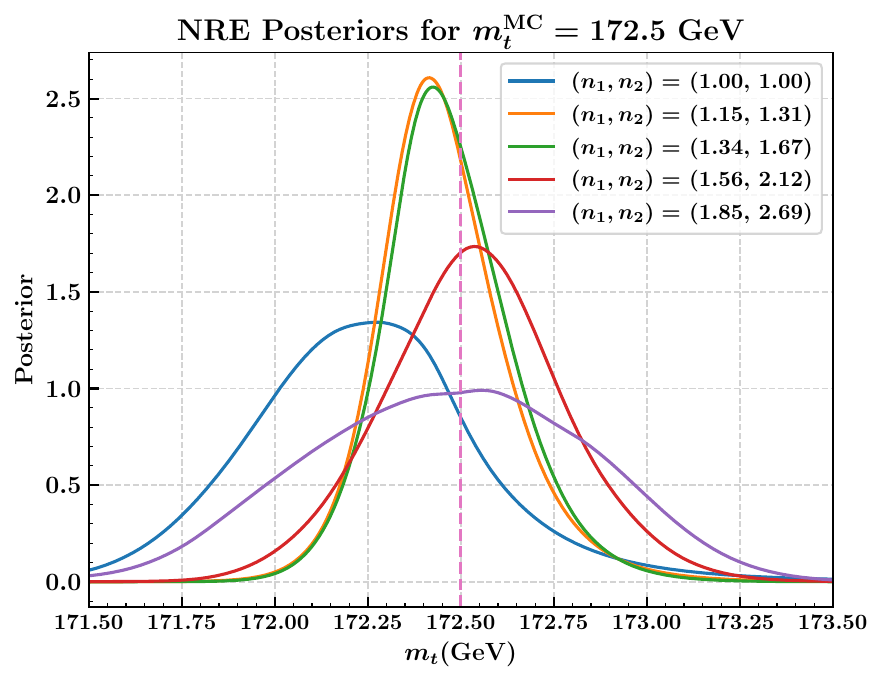} 
\label{fig:1725_posterior} 
} 
\subfloat[]{ \includegraphics[width=0.5\textwidth]{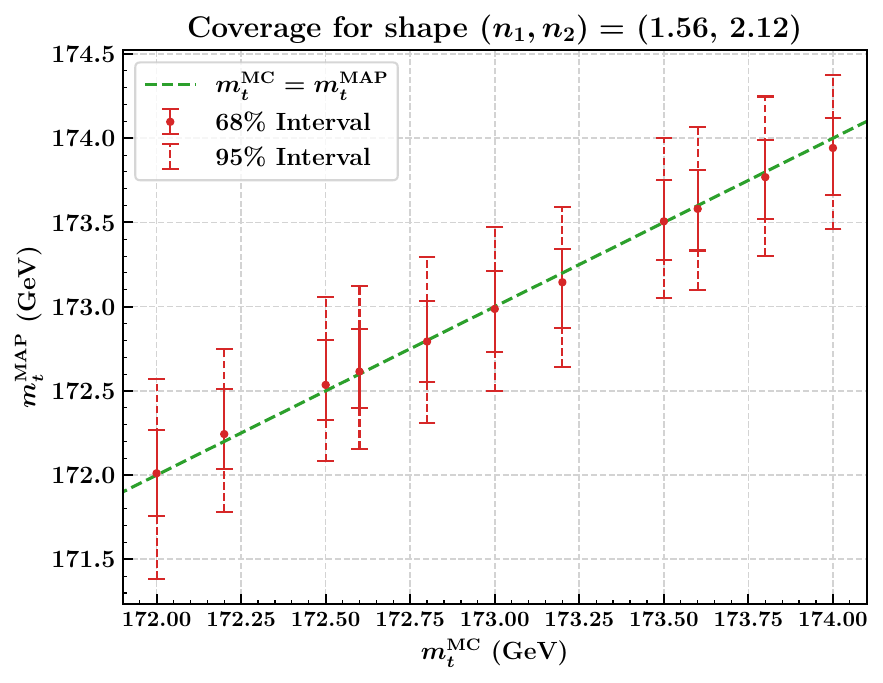}
\label{fig:coverage}
} 
\caption{Example posteriors from the NRE classifier. Left: Normalized posteriors for $\mtmc=172.5$ GeV for various shapes. Right: Uncertainty for the shape $(n_{1}, n_{2})=(1.56,2.12)$ with $\delta_{a}=0.01$ as a function of the Monte Carlo mass $\mtmc$. } 
\label{fig:nre_compare} 
\end{figure}

We consider three different metrics to determine the quality of the shape from the posterior. These are the maximum-a-posteriori $m_t^{\text{MAP}}$ (the mode of the posterior), and two different highest posterior density (HPD) intervals of the posterior. We associate the MAP to the extracted value of the top-quark mass and consider the HPD interval to be an uncertainty on this value. We remind the reader that the HPD interval associated with the value $k$ is defined as 
\begin{align}
    \text{HPD}_{k}=\{m_t~\vert~ p(m_t\vert \vec{\zeta}_{\text{Shape}}) \geq k\} 
\end{align}
which allows one to define a coverage or confidence interval 
\begin{align}
    \Theta_{\alpha} = \text{HPD}_{k}~ \text{s.t.}~  \int_{\text{HPD}_{k}}dm_{t}~p(m_{t}\vert\vec{\zeta}_{\text{Shape}})=\alpha\ .
\end{align}
We use $\alpha = 68\%,~95\%$. We show an example of the obtained posteriors for $\mtmc=172.5$ GeV for several different shapes in Fig~\ref{fig:1725_posterior} and the MAP and HPD intervals for one of them in Fig.~\ref{fig:coverage}. 

\begin{figure}[!htbp] 
\centering
    \subfloat[]{
        \includegraphics[width=0.48\textwidth]{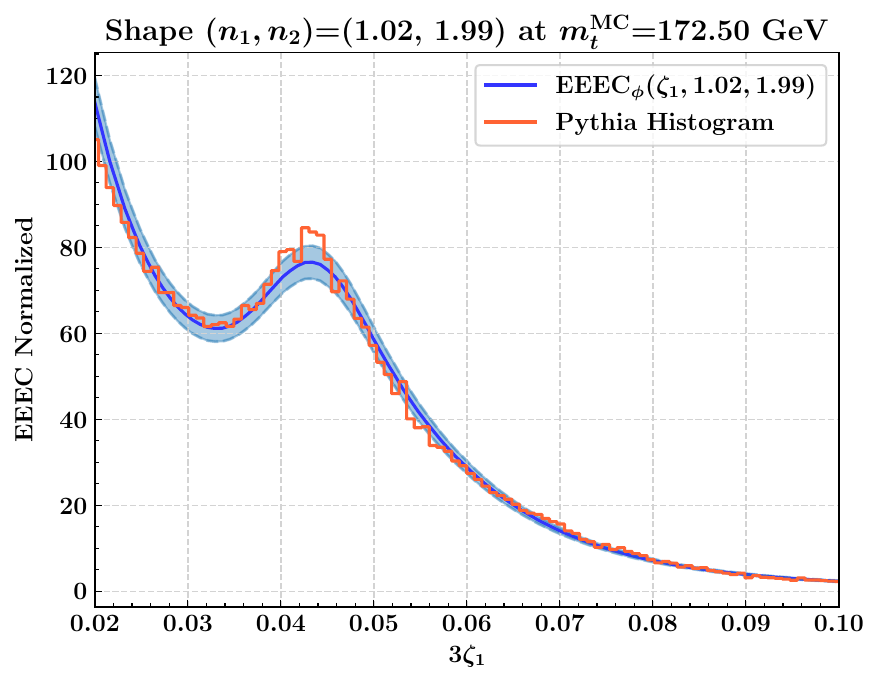}
        \label{fig:1725_posterior_opt}
    }
    \hfill
    \subfloat[]{
        \includegraphics[width=0.48\textwidth]{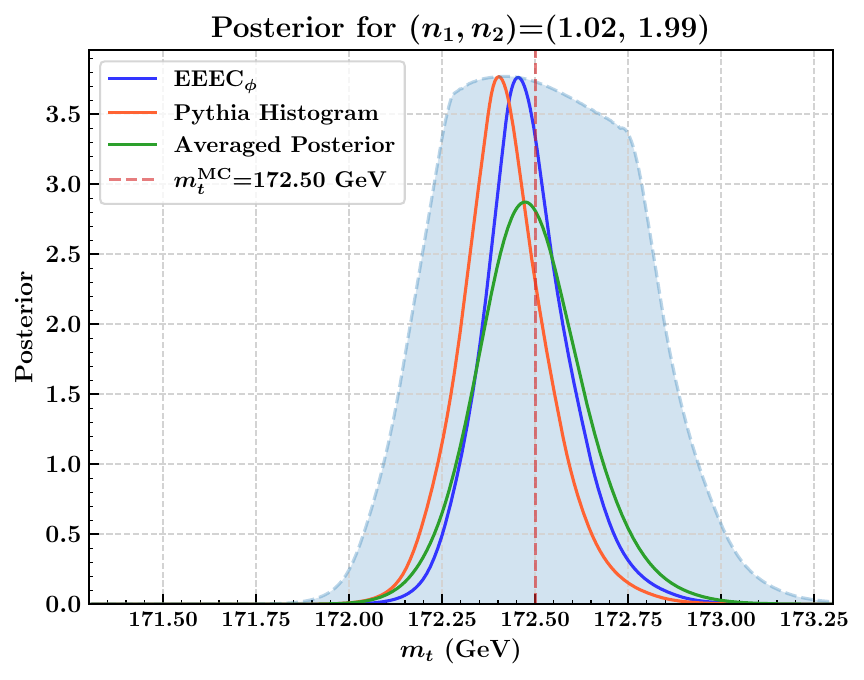}
        \label{fig:post_var}
    }
\caption{Comparison of posteriors computed directly on \textsc{Pythia} data to those from the DNN. Left: Comparison of EEEC with $(n_1, n_2)=(1.02,1.99)$ computed from the DNN with 5\% noise interval to direct computation on \textsc{Pythia} data. 
Right: Example posteriors, including a single DNN sample with $5\%$ Gaussian noise injected (dark blue), evaluation of the NRE network directly on the \textsc{Pythia} histogram (orange), and the envelope (shaded blue) and average (green) of 250 instances of the same DNN shape with different random $5\%$ noise. 
} 
\label{fig:optimal_shape} 
\end{figure}

In order to check the reliability of these metrics, we compute posteriors for various shape parameters $(n_1, n_2)$ with the Monte Carlo top mass $m_t^{\text{MC}}$ ranging over $[172, 174]$ GeV in steps of $0.25$ GeV using shapes extracted for the DNN and check that $m_t^{\text{MC}}$ is within the coverage interval. We generally find good agreement, typically with better agreement for the $95\%$ interval than for the $68\%$ one. We also confirm the reliability of these metrics by computing the posterior obtained by passing the \textsc{Pythia} histograms directly as input to the summary network during evaluation, finding that the resultant posterior lies well within the envelope of the allowed posteriors once the 5\% statistical noise is included for the DNN shapes.\footnote{Five percent is chosen due to being the approximate size of the variation of the \textsc{pythia} data from the DNN prediction for different samples. We also tested the network with less statistical error added to the DNN prediction and this shrinks the envelope of 250 replicas as expected, though the improvement plateaus before the noiseless limit is reached due to difficulty training as the amount of variability between samples shrinks.}
This is shown in Fig~\ref{fig:optimal_shape}, with the \textsc{pythia} histogram and DNN shape with a 5\% noise band for $(m_t^{\mathrm{MC}}, n_1, n_2) = (172.5, 1.02, 1.99)$ shown on the left in Fig~\ref{fig:1725_posterior_opt}, and the resultant posteriors on $m_t$ shown on the right in Fig~\ref{fig:post_var}, including the \textsc{pythia} histogram, a single DNN shape with 5\% noise, and the envelope and average of 250 DNN replicas.

\begin{figure}[!tbp]
    \centering
    \includegraphics[width=\columnwidth]{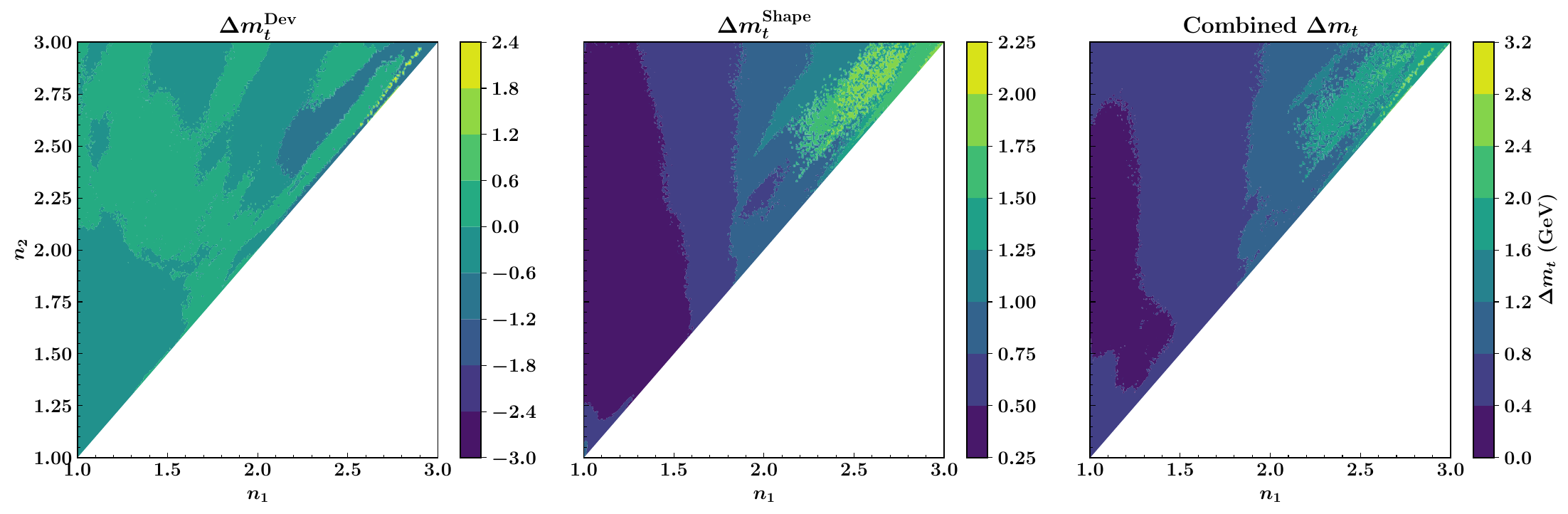}
    \caption{Error associated with EEEC shapes for $m_{t}^{\mathrm{MC}}=172.5$ GeV. For left to right, we plot the bias $\Delta m_{t}^{\mathrm{Dev}}$, variance $\Delta m_{t}^{\mathrm{Shape}}$, and the combined RMS average of the two quantities as a function of $(n_{1}, n_{2})$, with $n_{2}\geq n_{1}$.
    The diagonal line are isosceles triangles with one short and two long sides, while the left vertical line are isosceles triangles with one long and two short sides. Equilateral triangles are in the bottom left.
    }
    \label{fig:grid_search}
\end{figure}

Thus, having established that the NRE classifier produces reliable posterior estimates given an EEEC shape, we proceed to perform a search for an optimal shape that is maximally sensitive to the underlying $m_{t}^{\mathrm{MC}}$. The metric for estimating this sensitivity can be quantified using two quantities, as defined below
\begin{align}
    \Delta m_{t}^{\mathrm{Dev}} &= |m_{t}^{\mathrm{MAP}}-m_{t}^{\mathrm{MC}}| & \Delta m_{t}^{\mathrm{Shape}} &= \frac{\max{\Theta_{\alpha=0.95}}-\min{\Theta_{\alpha=0.95}}}{2}\ .
\end{align}
$\Delta m_{t}^{\mathrm{Dev}}$ and $\Delta m_{t}^{\mathrm{Shape}}$ quantify the bias and variance in the NRE posterior for a given shape. Assuming that these quantities are independent, we associate their root mean squared (RMS) average as the overall error for a given shape.\footnote{While it is unusual to add a bias and variance, both give us an important measure of how far the top-quark mass is from the true value. Additionally, other attempts to combine the two, such as taking the distance from the true $\mtmc$ to the furthest value within $1 \sigma$ of the posterior, are significantly more volatile.} The exercise is then to find the shape that minimizes it. We compute these quantities (by averaging over a set of 200 noisy shapes) for $(n_1, n_2)$ in the grid $[1,3]\times [1,3]$ consisting of 250 uniformly spaced points in each dimension, while enforcing the constraint $n_{2}\geq n_{1}$. The resultant errors on the grid can be computed for each $\mtmc$. For example, the error for $m_{t}^{\mathrm{MC}}=172.5$ GeV is shown in Fig.~\ref{fig:grid_search}. We find that the shape with the least error has 
\begin{align}
    (n_{1}^{\star}, n_{2}^{\star}) &= (1.02, 1.99) 
\label{eq:best_shape}
\end{align}
producing the estimate 
\begin{align}
    m_{t}^{\text{Inferred}} = 172.47_{\text{MAP}} \pm 0.31_{\Delta m_{t}^{\mathrm{Shape}}}\ \text{GeV}
\end{align}

Thus, the grid search finds a shape \eqref{eq:best_shape} where the triangle is nearly isosceles in the small angle limit, with the ratio of largest to smallest side being approximately $ \sqrt{n_{2}^{\star}} \sim \sqrt{2}$, (i.e) right isosceles triangles. While we expect our method for shape comparison to be robust, the actual number corresponding to this error should be taken with a grain of salt. The error here is primarily statistical, due to the statistical variation added to the density estimator during NRE training, and reflects the real statistical variation between shapes from data. That statistical noise is the source of error can be confirmed by seeing that the NRE error decreases when less noise is used during NRE training. Another potential source of error is the approximation of the DNN to the true EEEC distribution. We studied this in Fig.~\ref{fig:post_var} by evaluating the NRE classifier, which was trained using the surrogate model, on \textsc{pythia} data, finding that the difference between the surrogate model and data is not an important source of error because the posterior for the true simulation is well within the envelope of posteriors for the DNN. In order to include other sources of error such as systematic errors which are represented by varying Monte Carlo parameters, an additional study explicitly including MC variations would need to be performed. One might also want to understand the contribution to the error from other sources, such as from the neural network itself. While important, there is currently no consensus on quantifying these sources of error and studying them is a significant open question in the field (see \cite{Bhimji:2024bcd, Benato:2025rgo,Khot:2025kqg,Elsharkawy:2025six,Ghosh:2021roe,Nachman:2019dol,Gambhir:2022gua,Bollweg:2019skg,Bellagente:2021yyh, Rover:2024pvr, Bieringer:2024nbc, Bahl:2024gyt,Benevedes:2025nzr} for examples).

In order to ascertain if the uncertainty from the NRE posterior accurately reflects the inherent uncertainty in extracting the top mass from the EEEC shape, we compare them to estimates of the top mass from a classical fitting method. Our classical methodology is a synthesis of the methods expounded in refs.~\cite{Holguin:2022epo, Holguin:2024tkz, Flesher:2020kuy}. We first determine the peak of \textsc{Pythia} EEEC histograms. In order to do so, we fit degree 15 polynomials of the normalized EEEC as a function of the angular scale $\zeta_1$, and thereby compute analytically the position of the peak $\zeta_{\text{peak}}$. The process described above obtains a stable peak for polynomials of degree ranging from 12 to 17. Thereafter, we assume that the peak position $\zeta_\text{peak}$ is a linear function of $\left(m_t^{\text{MC}}/Q\right)^2$, and determine the corresponding $m_{t}^{\text{fit}}$ for a given $\zeta_{\text{peak}}$. To assign a statistical uncertainty to this classical fit, we obtain the EEEC histograms by using only random subsets of the data, retaining only a random 50$\%$ of all $\eec$ triplets from the 1M events we generated at each mass. This induces considerable jitter of the peak in each individual histogram, letting us extract an error on this polynomial fit from the statistical variation. For more details, we refer the interested reader to Appendix.~\ref{app:Classical_Fits}. We find that for $\mtmc=172.5$ GeV, the outlined classical method provides an estimate  for the optimal shape of $(n_1, n_2) = (1.02, 1.99)$
\begin{align}
m_{t}^{\mathrm{Fit}}=172.62 \pm 0.28\ \rm{GeV},
\end{align}
which gives an uncertainty of the same order of magnitude as that of the NRE estimate. This further validates our claim that the error extracted from the NRE is primarily statistical. However, the correlation between the error for this classical fit and the error extracted from the NRE is not perfect for all shapes, leaving room for future studies to understand these small differences. Additionally, in Appendix~\ref{app:Classical_Fits} we also demonstrate that if one was to use a high statistics test set from \textsc{Pythia}, and regress the top mass using a linear fit to the peak, the extracted $m_{t}^{\rm Fit}$ sits well within the confidence intervals from the NRE. 
Thus, we believe that the NRE analysis does allow a reasonable estimate of the `goodness' of the shape for top-quark mass regression.

\section{Conclusions \label{sec:conc}}

Modern machine learning has become ubiquitous in high energy physics, being used for tasks including but not limited to quark-gluon jet discrimination, boosted object identification, simulation, event reconstruction and parameter inference. For parameter inference as it relates to precision measurement in particular, the effective use of ML requires search methods which are limited to parameter space that is known to be theoretically calculable. In this paper, we perform such a search in the space of multi-dimensional energy-correlators, specifically for use extracting the top-quark mass. Being multi-dimensional in nature, a direct theory-based search is computationally prohibitive, and we use a two step ML approach to explore the observable space and find the optimal observable that is maximally sensitive to the mass. As the first step, we learn the underlying 3 point energy correlator distribution using both a dense neural network with a novel physics-motivated loss function and a normalizing flow. As the second step, we used the dense network as a surrogate model to rapidly produce EEECs of varying shapes, and use neural ratio estimation to compute the posterior on the underlying top-quark mass from the shapes. We then pick an optimal shape within the training dataset by minimizing the width of this posterior. This gives an observable which can then be directly compared between precision theory and experiment without ML input. Because the output of this process is an observable, bias or error in our NN search on Monte Carlo simulation may prevent our selected observable from being optimal when applied to actual data, but will not bias the final measurement. Our search is limited to a subset of one dimensional observable parameterizations, but it would also be interesting to study the full distribution in more detail to understand the limit of how well these marginalizations can perform. More generally, studying the full distribution of energy correlators in other contexts might give us more insight about their structure.

While we only demonstrated this approach for EEECs in the specific case of learning the top-quark mass, we expect that the outlined ML approach works for optimizing a large class of precision collider observables for parameter inference. This type of observable optimization could serve as an alternative to needing to understand uncertainty estimation in NNs directly (an interesting and growing field, see \cite{Bhimji:2024bcd, Benato:2025rgo,Khot:2025kqg,Elsharkawy:2025six,Ghosh:2021roe,Nachman:2019dol,Gambhir:2022gua,Bollweg:2019skg,Bellagente:2021yyh, Rover:2024pvr, Bieringer:2024nbc, Bahl:2024gyt,Benevedes:2025nzr}). In addition, these techniques can complement other methods of designing understandable observables \cite{Long:2023mrj, Bahl:2025jtk} or could be used to help understand the  optimal reduction of higher dimensional observables to lower dimensional ones more generally.


\section{Acknowledgements}
The authors thank Aur\'elien Dersy, Jesse Thaler, Rikab Gambhir, Ben Nachman, Vinicius Mikuni, Matthew Reece, Dennis Noll, Sascha Diefenbachar, and David Shih for useful discussions. AB and MDS are supported by DOE grant DE-SC0013607. KF is supported in part by: DOE grant DE-SC0013607, the Harvard GSAS Merit Fellowship, and the Miller Institute for Basic Research in Science, University of California Berkeley. This work is also supported by the National Science Foundation under Cooperative Agreement PHY-2019786 (The NSF AI Institute for Artificial Intelligence and Fundamental Interactions, \href{http://iaifi.org/}{http://iaifi.org/}). KF also thanks the The Munich Institute for Astro-, Particle and BioPhysics and the Aspen Center for Physics (which is supported by NSF grant PHY-2210452, Simons Foundation grant (1161654, Troyer), and Alfred P. Sloan Foundation grant G-2024-22395) for hospitality while working on this project. The computations in this paper were performed on the Harvard Cannon Cluster, including resources provided by the Institute for Artificial Intelligence and Fundamental Interactions (IAIFI).

\appendix

\section{Classical Fits}\label{app:Classical_Fits}

\begin{figure}[!htb]
    \centering
    \includegraphics[scale=0.6]{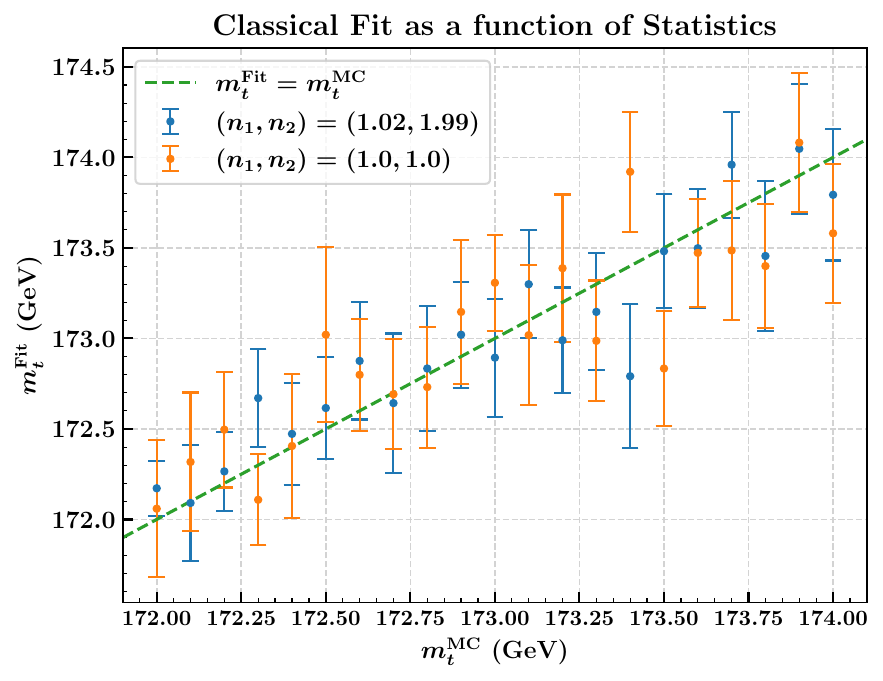}
    \caption{Comparison of the classically extracted $m_{t}^{\rm Fit}$ for the shapes $(n_1,n_2)_{\delta_{a}}=(1.02,1.99)_{0.01}$, and $(n_1,n_2)_{\delta_{a}}=(1.0,1.0)_{0.01}$ extracted using a polynomial fit to the peak. The statistical spread arises from a random selection of the EEEC tuples, extracting the peak $\zeta_{\rm peak}$ of the shape from the random set, and thereafter performing a linear fit $\zeta_{\rm peak}$ as a function of $\mtmc$ using only the median peak data. It is evident that there is a sizable uncertainty in the extracted top masses due to limited statistics near the peak region.
    }
    \label{fig:classical_fit}
\end{figure}

In this appendix, we elaborate on the method we use to extract the top-quark mass from an $\eec$ marginal (see e.g., Eq.~\ref{eqn:shape}) without a NN. As noted in references~\cite{Holguin:2022epo,Holguin:2023bjf,Holguin:2024tkz}, the peak of the marginalized $\eec$ is linearly dependent on the square of the top mass $\mtmc$, i.e. 
\begin{align}
    \zeta_{\rm peak} &= a \left(\frac{\mtmc}{Q}\right)^2 + b \label{eq:lin_peak}
\end{align}
where $a,b$ depend on the shape parameters $n_{1},n_{2},\delta_{a}$. For example, \cite{Holguin:2022epo} found that for unclustered jets, with equilateral triangles ($n_{1}=1,n_{2}=1,\delta_{a}=0.02$), $a\approx 3$ and $b\approx 0$. For the shapes we consider, we extract $a$, $b$ by performing a linear of the form in Eq.~\eqref{eq:lin_peak} across different $m_t^{\mathrm{MC}}$ values, and then use the best-fit parameters to solve for $m_t$ for a given peak, i.e. 
\begin{align}
    m_{t}^{\rm Fit} = Q \sqrt{\frac{\zeta_{\rm peak}-b_{\rm Fit}}{a_{\rm Fit}}} \label{eq:invert}
\end{align}
For this fit, we use $\mtmc \in [172,174]$ GeV, with step size of $0.1$ GeV.

While this procedure seems straightforward, its reliability relies on the the value of the peak $\zeta_{\mathrm{peak}}$ in a given data sample being a reliable proxy for the true $\zeta_{\mathrm{peak}}$ at a given $m_t^{\mathrm{MC}}$. However, because the $\eec$ is an ensemble observable where a small number of tuples contribute a large amount to the peak even for substantial numbers of jets, there is significant statistical uncertainty in the location of $\zeta_{\rm peak}$ for a given data sample. This jitter in the peak position will translate into a sizable uncertainty in the regressed $m_{t}^{\rm Fit}$. To understand the approximate size of this error, we considered 20 different randomly selected subsamples (with replacement) for each $\mtmc$, each containing half of the total triplets from 1M jets.

\begin{figure}[h] 
\centering 
\subfloat[]{\includegraphics[width=0.5\textwidth, height=5.6cm, keepaspectratio]{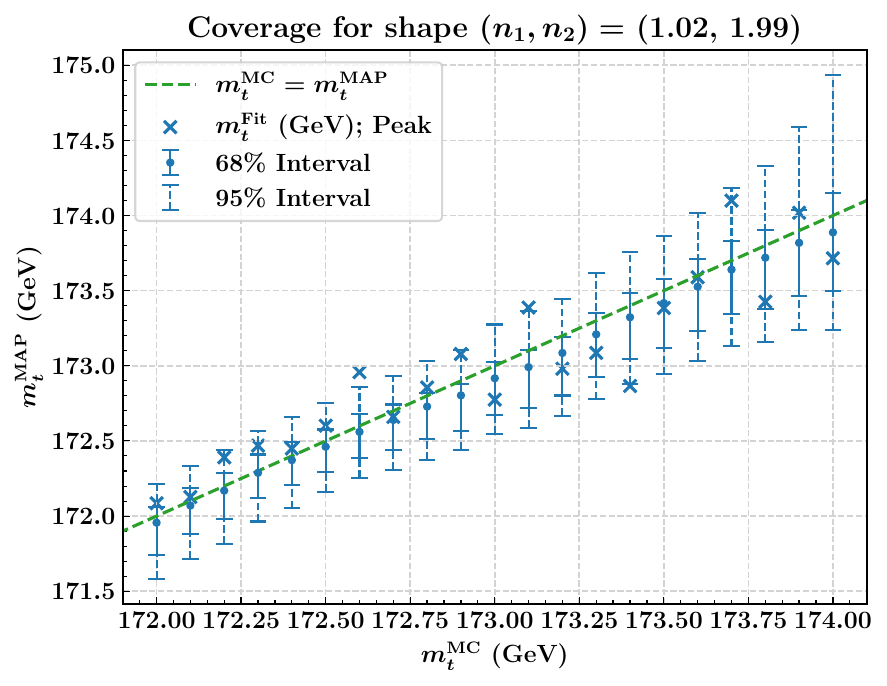} 
\label{fig:cov_best_shape} 
} 
\hfill
\subfloat[]{ \includegraphics[width=0.5\textwidth, height=5.6cm, keepaspectratio]{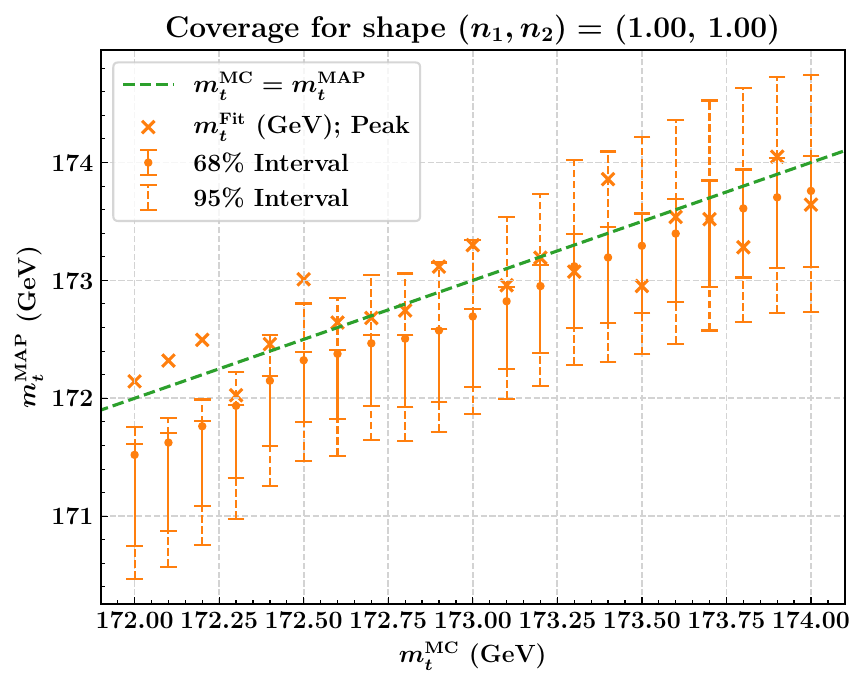}
\label{fig:cov_equi}
}
\caption{Comparison of the peak extracted from the classical polynomial fit to the corresponding NRE estimates for multiple different shapes. Left: $(n_1, n_2) = (1.02, 1.99)$. Right: Equilateral triangles with $(n_1, n_2) = (1.0, 1.0)$. We find that the $95\%$ confidence intervals from the NRE encompass the extracted values of $m_{t}^{\rm Fit}$ from classical peak fitting, thereby indicating that the NRE uncertainties are reflective of underlying uncertainty of each shape. 
} 
\label{fig:cov_best} 
\end{figure}

 For each, we computed the corresponding $\eec$ restricted such that for the smallest side $\zeta_1$, $3\zeta_1\in[0.02,0.1]$, and then fit a 15 degree polynomial to the histogram. We analytically found the local maximum for each and use the median $\zeta_{\mathrm{peak}}$ at each $\mtmc$ to obtain the best fit $a$ and $b$ with Eq.~\eqref{eq:lin_peak}. We then compute the standard deviation in $m_t$ values extracted using Eq.~\eqref{eq:invert} and call this the statistical error for the linear fit. An example is shown in Fig.~\ref{fig:classical_fit} for shapes $(n_1,n_2)_{\delta_{a}}=(1.02,1.99)_{0.01}$, and $(n_1,n_2)_{\delta_{a}}=(1.0,1.0)_{0.01}$. We find that 
 the polynomial fit outlined above does yield errors that are of the same magnitude as that from the NRE, showing that statistical uncertainty dominates both the NRE and polynomial fit predictions for $m_t$. Because of this, we expect the NRE to be a relatively reliable proxy for the classical error, even though the correlation between the two is imperfect and is an interesting direction for future work. Additionally, we also check that the $\mtmc$ value extracted using a linear fit with higher statistics (from full set of triplets from 1M events) typically lies within the NRE confidence interval, further improving our confidence in the NRE predictions. This is shown in Fig.~\ref{fig:cov_best}.

\newpage
\bibliographystyle{utphys}
\bibliography{references.bib}
\end{document}

%% file: Tikz/mlp_styles.tikzstyles
\tikzstyle{inputbox}=[fill={rgb,255: red,255; green,253; blue,222}, draw=black, shape=rectangle]
\tikzstyle{mlpnode}=[fill={rgb,255: red,203; green,255; blue,252}, draw=black, shape=circle]
\tikzstyle{noise box}=[fill={rgb,255: red,221; green,228; blue,255}, draw=black, shape=rectangle]

\tikzstyle{edge style 0}=[->]
\tikzstyle{dashes}=[-, dash pattern=on 2pt off 1pt, fill={rgb,255: red,255; green,223; blue,217}]

%% file: intro.tex
\section{Introduction}\label{sec:intro}

There are many observables in collider physics which are  useful  but essentially impossible to calculate from first principles. For example, the output of a neural network trained on simulation to distinguish quark from gluon jets may have phenomenal discrimination power, but will never be computable without a simulation. There are many other observables which can be computed to high precision, such as the 8-point gluon-scattering amplitude in $\mathcal{N}=4$ Super-Yang Mills theory, but offer little hope of ever being measured. Unfortunately, the intersection space of systematically calculable observables and useful observables is an infinitesimal subset of all observables. In this paper, we suggest that this intersection space may be fruitfully explored using machine learning. The basic idea is to search for points within a multi-dimensional space of calculable observables that are optimal for some task. We do this by first learning the multidimensional distribution and then searching within that space using a neural network. The output is an observable which can subsequently be calculated and measured without further reference to the neural network which suggested it.


To explore the space of computable and useful observables for precision collider physics, our approach is first to select some manageable subset of all observables which are
in principle calculable. For example, we could consider jet rates, or event shapes, or number of leptons. In this paper we focus on energy-energy correlators. For any low-dimensional space of in-principle calculable observables, we would like to narrow in on which point in this space is most useful for some task. Critically, this selection can be done before the precision computation is performed. To this end, we employ Monte-Carlo (MC) simulations to sample very high-dimensional kinematic distributions which can then be marginalized down to the subspace of interest~\cite{Bierlich:2022pfr,Buckley:2011ms,Bahr:2008pv,Gleisberg:2003xi,Alwall:2014hca}.
The next step is to search within this low-dimensional space for a point of maximum
`sensitivity' 
to underlying physical parameters of interest such as coupling strengths and particle masses. 
Although it is challenging to search in the space of observables directly using the Monte-Carlo simulated data, the search can be efficiently done by using modern machine learning  techniques (ML) to first learn the distribution in the low-dimensional theory-aware subspace, then to search within that space of observables. The result is then an observable which can be measured directly on data and compared to precision theory without further reference to either the Monte Carlo simulation or to the machine learning. 
A caricature of the strategy is shown in Fig.~\ref{fig:flow}.

{
\input{Tikz/ml_flow.tikzstyles}
\begin{figure}
    \centering
    \scalebox{1.15}{\input{Tikz/workflow.tikz}}
    \caption{Schematic of the ML workflow which can explore the space of precision-compatible observables. One first learns an analytic surrogate of the multidimensional observable as a function of physical parameters such as couplings and masses, followed by solving the inverse problem of inferring those parameters given marginals of or uni-dimensional observables from those distributions. The latter step is where observable quality can be quantified by the fidelity of the inferred physical parameters.}
    \label{fig:flow}
\end{figure}
}
The task of estimating a complex high-dimensional distribution given only its samples (from a simulator or experiment) is a difficult one, if done solely using numeric statistical methods such as kernel density estimation \cite{10.1214/aoms/1177728190,10.1214/aoms/1177704472}. With the advent of better computing, machine learning has emerged as a readily adaptable approach for probability density function (PDF) estimation (e.g \cite{Louppe:2017ipp, Andreassen:2018apy, Andreassen:2019nnm}). 
Many different modern ML architectures which minimize the negative log likelihood loss have been developed for density estimation \cite{Andreassen:2018apy, Andreassen:2019txo, Papamakarios:2017tec, Sengupta:2023xqy}. Simple networks such as dense neural networks (DNNs) can be used for density estimation given the proper loss. More elaborate networks such as \emph{normalizing flows} and their continuous variants~\cite{Dinh:2014mzt, Dinh:2016pgf, Rezende:2015,Papamakarios:2019,Chen:2018,Song:2019,Dockhorn:2022, Grathwohl:2018vdm} are also commonly used. 
By learning an invertible map from data distributions to simpler base distributions such as Gaussians, normalizing flows (NF) produce an analytic estimate of the underlying PDF. NFs have gained popularity in collider physics, and have been used for wide ranging applications including anomaly detection~\cite{Golling:2022nkl, Krause:2023uww, Hallin:2021wme,Butter:2022lkf,Das:2024fwo,Nachman:2020lpy,Hallin:2022eoq,Das:2023bcj}, calibration~\cite{Du:2024gbp}, decorrelation~\cite{Klein:2022hdv}, detector simulation~\cite{Mikuni:2022xry,Ernst:2023qvn,Du:2024gbp,Pang:2023wfx,Diefenbacher:2023flw, Dreyer:2025zhp, Buss:2024orz, Diefenbacher:2023vsw, Gao:2020zvv, Krause:2021wez,Krause:2021ilc,Krause:2022jna,Buckley:2023daw,Favaro:2024rle}, phase space integration~\cite{Gao:2020vdv, Heimel:2022wyj,Heimel:2023ngj}, reconstruction~\cite{Raine:2023fko, Leigh:2022lpn},  reweighting~\cite{Golling:2023mqx, Algren:2023qnb}, and unfolding~\cite{Butter:2024vbx,Chan:2023tbf, Buhmann:2023zgc}. While a straightforward application of density estimation techniques to learning observable distributions seems obvious, in some cases we find that physics-inspired loss functions which learn a weighted distribution rather than the probability perform better. Such an approach quantitatively weights the network attention to learning the pertinent parts of the PDF (which may be sparsely populated) that are sensitive to the underlying physics parameters of interest.

Having constructed parametrized observable distributions, one moves onto the second step of observable space exploration - to characterize their sensitivity to underlying physical parameters. Here one needs to solve the inverse problem of inferring the causal physics parameters given an observable distribution. ML again proves to be an incredibly useful tool for parameter inference by providing methods that are broadly termed as \emph{neural simulation-based inference} (NSBI) \cite{Cranmer:2015bka,Papamakarios:2016,Papamakarios:2018zoy,Cranmer:2019eaq,Brehmer:2020cvb,Zammit-Mangion:2024}. NSBI methods provide accurate computation of several statistical quantities of importance, such as the joint and marginal likelihoods from distributions which enable a quantitative metric on the inferred physical parameters. NSBI methods are also advantageous compared to traditional Bayesian computation methods as they amortize the inference process, allowing priors to be changed during inference without retraining the relevant networks. 
An example of NSBI that amortizes inference  
is \emph{neural ratio estimation} (NRE)~\cite{Hermans:2019,Cranmer:2015bka,Cole:2021gwr,Miller:2022haf}. NRE employs classifier neural networks that learn to discriminate between samples labeled with their true underlying parameters and samples with mislabeled parameters. The resultant classifier loss can be shown to approximates the ratio of the parameter posterior to prior, allowing one to quantitatively infer the most probable parameter that led to an observed sample.  
Several such NSBI methods have gained some popularity for parameter inference in astrophysical systems~\cite{Dax:2021tsq,Zhao:2021ddh,Legin:2021zup,Mishra-Sharma:2021oxe,Shih:2023jme,Schosser:2024aic, Liang:2025yjw, Coogan:2022cky, List:2023jwo,AnauMontel:2023stj,AnauMontel:2022ppb,Zeghal:2024kic}, and have also been previously used in particle physics in other contexts \cite{Andreassen:2019nnm, Andreassen:2020gtw, Flesher:2020kuy, Acosta:2025lsu, Ghosh:2025fma, ATLAS:2025clx, Shyamsundar:2025eht, Cheng:2025ewj, Sluijter:2025isc, Bahl:2024meb}.  

In this paper we present a concrete proof-of-principle application of these ideas: we search in the space of energy-energy correlators for a point which is optimally sensitive to the top-quark mass. 
EECs~\cite{Basham:1977iq,Basham:1978bw,Basham:1978zq,Basham:1979gh,Korchemsky:1997sy,Berger:2003iw} have recently seen a revival as a calculable precision observable in high energy collider physics~\cite{Bauer:2008dt,Larkoski:2013eya,Moult:2016cvt,Dixon:2019uzg,Chen:2020vvp,Gao:2019ojf,Komiske:2022enw,Chen:2022swd,Lee:2022uwt,Jaarsma:2024ngl,Lee:2023npz,Alipour-fard:2024szj,Lee:2024esz}; see \cite{Moult:2025nhu} for a recent review. EECs describe the angular correlation and distribution of the energy flow arising from particle collisions at null future infinity, thereby serving as a field-theoretic definition of calorimeters
~\cite{Sveshnikov:1995vi,Cherzor:1997ak,Tkachov:1995kk}. Being integrated functions of the stress-energy tensor in QFT, they have allowed significant import of insights from conformal field theories to precision collider physics~\cite{Hofman:2008ar,Belitsky:2013xxa,Belitsky:2013bja,Kravchuk:2018htv,Chen:2022jhb,Chen:2021gdk,Csaki:2024joe}. Phenomenologically, they have enabled a novel exploration of jet substructure analysis at hadron and lepton colliders \cite{CMS:2024mlf,Komiske:2022enw},
precision measurements of heavy quark dynamics such as the dead cone effect~\cite{Gudima:1979vj,Armesto:2003jh,Craft:2022kdo,Aglietti:2024zhg}, and provided a bridge for interpolating between observables for proton collisions and heavy ion physics~\cite{Andres:2022ovj,Andres:2023xwr,Devereaux:2023vjz,Andres:2023ymw,Rai:2024ssx,Barata:2024bmx,Andres:2024ksi,Barata:2024wsu,Liu:2023aqb}.

The general $k$-point EEC distribution is schematically given by 
\begin{align}
\langle E(\vec{n}_1) E(\vec{n}_2) \cdots E(\vec{n}_k)\rangle
    = \frac{1}{\sigma}\int d\sigma \times E(\vec{n}_1)\times E(\vec{n}_{2}) \times\cdots \times E(\vec{n}_k)\ . 
\end{align}
where $\{\vec{n}\}_{i=1}^{k}$ denote the $k$ directions of energy flow. The $k$-point EEC thus weights the cross section for a process by the product of energy $\prod_{i=1}^{n}E(\vec{n}_{i})$ flowing in a set of prescribed directions.
Experimentally, the EEC is measured by recording the product of the energies for each $k$-particle subset of all particles in a collision at a point characterizing the relative angles between the particles. This measurement is then summed over both all the subsets of $k$-particles in each event and then over all events.
EECs characterize the pattern of energy flow arising from particle collisions as functions of the angular distances between directions. The energy weighting inherent in the definition of an EEC is extremeley useful, as it suppresses the contribution from soft radiation from pileup and the underlying event, which is under relatively poor theoretical control. In fact, compared to  observables that groom for decontaminating soft radiation such as soft drop jet mass, EECs have stood out by furnishing high-order distribution predictions for hadron colliders which have led to some precise measurements~\cite{ATLAS:2017qir,ATLAS:2023tgo,CMS:2024mlf}. 

As $k$ increases, $k$-point EECs capture in increasing detail the underlying physics at different energy scales. The simplest EEC observable, the two-point function, suffices to capture soft-collinear physics from QCD jets~\cite{Kardos:2018kqj,Gao:2019ojf,Gao:2023ivm}, and the two-pronged decays of the $W$ boson~\cite{Ricci:2022htc}. Analytic computations of multi-point EECs are currently limited to $3$ point in QCD~\cite{Yang:2022tgm} and $4$ point in $\mathcal{N}=4$ SYM \cite{Chicherin:2024ifn}. 
In this work, we focus on using EECs to measure the top-quark mass, following~\cite{Holguin:2022epo,Holguin:2023bjf, Holguin:2024tkz, Xiao:2024rol}. The top quark decays to three hard partons, so it is natural to search within the space of $3$-point EECs for sensitivity to the top-quark mass. In particular, we will look at highly energetic top quarks since the boost makes the three decay particles relatively collimated and the associated EEEC ($3$-point EEC) then becomes particularly well-suited for both measurement and precision theory. This was first considered in Ref.~\cite{Holguin:2022epo}, where the 1-dimension distribution of nearly equilateral-triangle three-point EECs was studied. More recent papers have looked at a couple of other shapes for three-point correlators~\cite{Holguin:2023bjf, Holguin:2024tkz, Xiao:2024rol}, finding increased performance compared to the equilateral case. This improvement motivates our systematic search for a marginalization which is optimally suited to top-quark mass measurement.

The rest of the paper is organized as follows. In Sec.~\ref{sec:eec_density} we describe our physics-motivated ML approaches to learning the EEEC distribution arising from boosted top jets in $e^+ e^-$ collisions. We detail the methods we use for density estimation, and show that marginalizing these learned distributions correctly reproduces different EEEC marginalizations of the data. Next, we show how NRE can be used on marginalizations of the full distribution to explore one-dimensional energy correlator space and extract the top-quark mass in Sec.~\ref{sec:fit}. Finally, we conclude in Sec.~\ref{sec:conc}.

%% file: Tikz/ml_flow.tikzstyles

\tikzstyle{sq_box_0}=[fill={rgb,255: red,203; green,246; blue,255}, draw=black, shape=rectangle]
\tikzstyle{circ_0}=[fill={rgb,255: red,255; green,191; blue,191}, draw={rgb,255: red,255; green,191; blue,191}, shape=circle]

\tikzstyle{arrow 0}=[->]
\tikzstyle{edge_col_0}=[-, fill={rgb,255: red,255; green,205; blue,255}]

%% file: Tikz/workflow.tikz
\begin{tikzpicture}
	\begin{pgfonlayer}{nodelayer}
		\node [style=none] (1) at (-3.75, 6.5) {};
		\node [style=none] (2) at (2.25, 6.5) {};
		\node [style=none] (3) at (-3.75, 3.5) {};
		\node [style=none] (4) at (2.25, 3.5) {};
		\node [style=none] (5) at (-0.75, 5) {$\vec{x}_{\mathrm{Obs}}(\alpha_s,m_{t},\cdots)$};
		\node [style=none] (6) at (-4.25, 5) {};
		\node [style=none] (7) at (-3.75, 2) {};
		\node [style=none] (8) at (2.25, 2) {};
		\node [style=none] (9) at (-3.75, -1) {};
		\node [style=none] (10) at (2.25, -1) {};
		\node [style=none] (11) at (-0.75, 0.5) {$p(m_{t},\alpha_{s},\cdots\vert \vec{x}_{\mathrm{Obs}})$};
		\node [style=none] (14) at (-4.25, 0.5) {};
		\node [style=none] (15) at (2.75, 5) {};
		\node [style=none] (16) at (2.75, 0.5) {};
		\node [style={sq_box_0}] (17) at (6, 2.5) {Inference};
		\node [style={sq_box_0}] (18) at (-7.25, 2.5) {$\mathcal{L}(\alpha_s, m_{t})$};
		\node [style=none] (19) at (-0.75, 6.5) {};
		\node [style=none] (20) at (-0.75, -1) {};
	\end{pgfonlayer}
	\begin{pgfonlayer}{edgelayer}
		\draw [style={edge_col_0}] (4.center)
			 to (3.center)
			 to [bend left] (1.center)
			 to (2.center)
			 to [bend left] cycle;
		\draw [style={edge_col_0}] (10.center)
			 to (9.center)
			 to [bend left] (7.center)
			 to (8.center)
			 to [bend left] cycle;
		\draw [style=arrow 0, bend left=90, looseness=1.50] (15.center) to (16.center);
		\draw [style=arrow 0, bend left=90, looseness=1.75] (14.center) to (6.center);
	\end{pgfonlayer}
    \node[anchor=south, xshift=0pt, yshift=0pt] at (19.north) {%
        \includegraphics[scale=0.35]{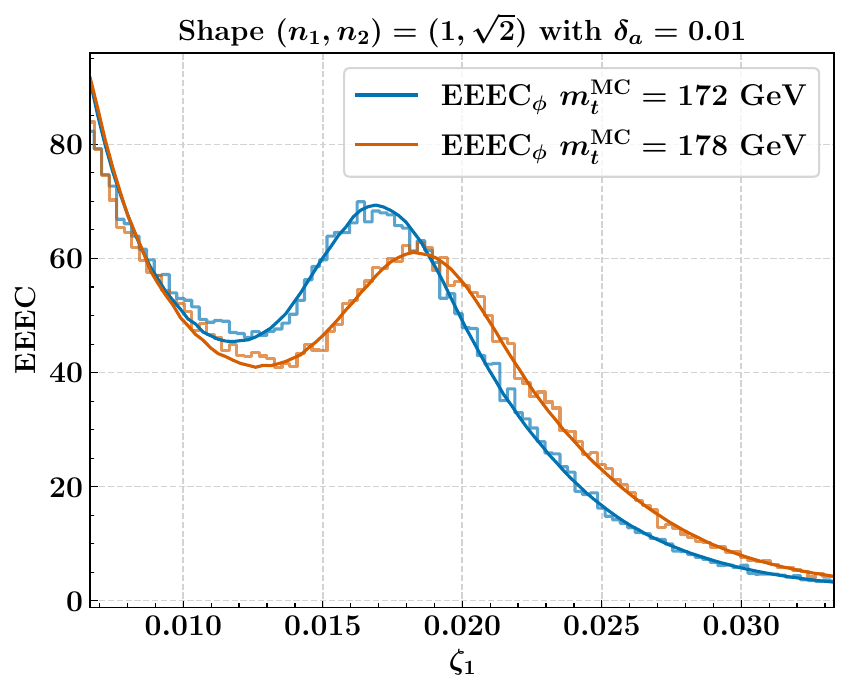}%
      };
        \node[anchor=north,
        xshift=0pt, yshift=0pt] at
        (20.south) {%
        \includegraphics[scale=0.35]{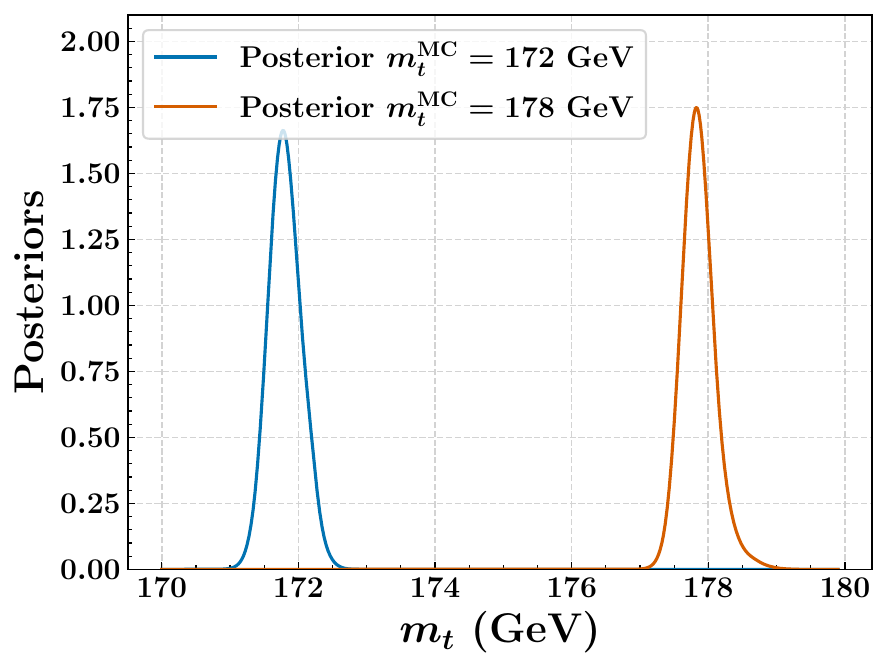}
        };
\end{tikzpicture}

%% file: Tikz/mlp_example.tikz
\begin{tikzpicture}
	\begin{pgfonlayer}{nodelayer}
        \node [style=inputbox,
  label={[yshift=-0.02cm]south:{\includegraphics[scale=0.1]{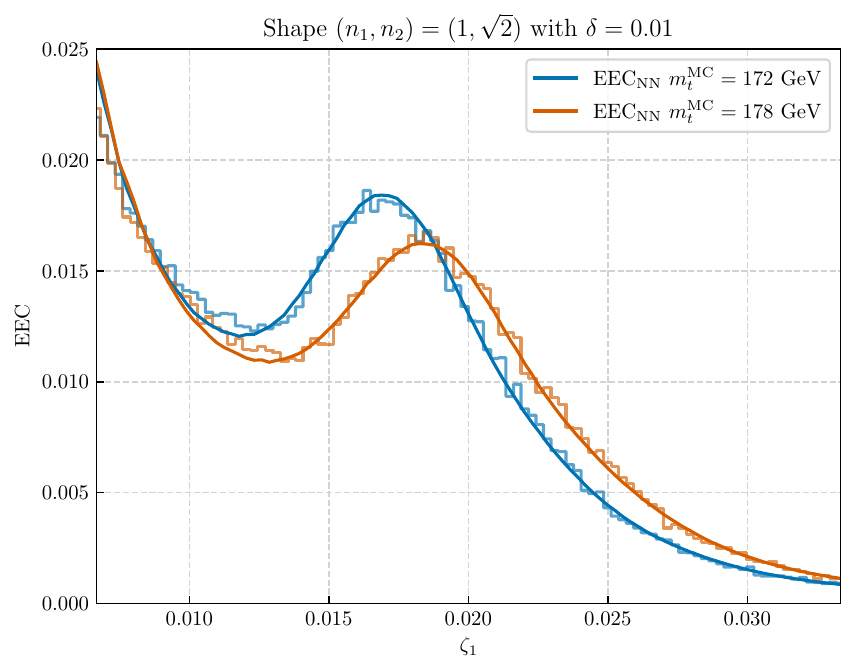}}}
] (0) at (-10,5) {$\vec{\zeta}_{\rm Shape}$};
		\node [style=inputbox] (1) at (0, 3) {$m_{t}$};
		\node [style=inputbox] (2) at (0, 1.5) {$n_{1}$};
		\node [style=inputbox] (3) at (0, 0) {$n_{2}$};
		\node [style=mlpnode] (4) at (-6, 6) {};
		\node [style=mlpnode] (5) at (-6, 5.25) {};
		\node [style=mlpnode] (6) at (-6, 4.5) {};
		\node [style=mlpnode] (7) at (-6, 3.75) {};
		\node [style=none] (8) at (-5, 6) {};
		\node [style=none] (9) at (-5, 5.25) {};
		\node [style=none] (10) at (-5, 4.5) {};
		\node [style=none] (11) at (-5, 3.75) {};
		\node [style=mlpnode] (12) at (-5, 6) {};
		\node [style=mlpnode] (13) at (-5, 5.25) {};
		\node [style=mlpnode] (14) at (-5, 4.5) {};
		\node [style=mlpnode] (15) at (-5, 3.75) {};
		\node [style=none] (16) at (-4, 6) {};
		\node [style=none] (17) at (-4, 5.25) {};
		\node [style=none] (18) at (-4, 4.5) {};
		\node [style=none] (19) at (-4, 3.75) {};
		\node [style=none] (24) at (-3, 6) {};
		\node [style=none] (25) at (-3, 5.25) {};
		\node [style=none] (26) at (-3, 4.5) {};
		\node [style=none] (27) at (-3, 3.75) {};
		\node [style=inputbox] (32) at (0.5, 5) {$s_{\phi}(\vec{\zeta}_{\rm shape})$};
		\node [style=mlpnode] (33) at (4, 2.75) {};
		\node [style=mlpnode] (34) at (4, 2) {};
		\node [style=mlpnode] (35) at (4, 1.25) {};
		\node [style=mlpnode] (36) at (4, 0.5) {};
		\node [style=none] (37) at (5, 2.75) {};
		\node [style=none] (38) at (5, 2) {};
		\node [style=none] (39) at (5, 1.25) {};
		\node [style=none] (40) at (5, 0.5) {};
		\node [style=mlpnode] (41) at (5, 2.75) {};
		\node [style=mlpnode] (42) at (5, 2) {};
		\node [style=mlpnode] (43) at (5, 1.25) {};
		\node [style=mlpnode] (44) at (5, 0.5) {};
		\node [style=none] (45) at (6, 2.75) {};
		\node [style=none] (46) at (6, 2) {};
		\node [style=none] (47) at (6, 1.25) {};
		\node [style=none] (48) at (6, 0.5) {};
		\node [style=none] (53) at (7, 2.75) {};
		\node [style=none] (54) at (7, 2) {};
		\node [style=none] (55) at (7, 1.25) {};
		\node [style=none] (56) at (7, 0.5) {};
		\node [style=inputbox] (61) at (11, 1.5) {$f_{\phi}$};
		\node [style=none] (62) at (-6.75, 6.75) {};
		\node [style=none] (63) at (-6.75, 3) {};
		\node [style=none] (64) at (-2.25, 6.75) {};
		\node [style=none] (65) at (-2.25, 3) {};
		\node [style=none] (66) at (3, 3.5) {};
		\node [style=none] (67) at (8, 3.5) {};
		\node [style=none] (68) at (8, -0.25) {};
		\node [style=none] (69) at (3, -0.25) {};
		\node [style=noise box] (70) at (-7.75, 6.5) {$\epsilon$};
		\node [style=none] (71) at (-7.75, 5) {};
		\node [style=none] (72) at (-4.5, 2.5) {$\rm Summary\ Network$};
		\node [style=none] (73) at (5.5, -0.75) {$\rm Classifier\ Network$};
		\node [style=mlpnode] (74) at (-5, 6) {};
		\node [style=mlpnode] (75) at (-5, 5.25) {};
		\node [style=mlpnode] (76) at (-5, 4.5) {};
		\node [style=mlpnode] (77) at (-5, 3.75) {};
		\node [style=none] (78) at (-4, 6) {};
		\node [style=none] (79) at (-4, 5.25) {};
		\node [style=none] (80) at (-4, 4.5) {};
		\node [style=none] (81) at (-4, 3.75) {};
		\node [style=mlpnode] (82) at (-4, 6) {};
		\node [style=mlpnode] (83) at (-4, 5.25) {};
		\node [style=mlpnode] (84) at (-4, 4.5) {};
		\node [style=mlpnode] (85) at (-4, 3.75) {};
		\node [style=none] (90) at (-3, 6) {};
		\node [style=none] (91) at (-3, 5.25) {};
		\node [style=none] (92) at (-3, 4.5) {};
		\node [style=none] (93) at (-3, 3.75) {};
		\node [style=mlpnode] (94) at (-3, 6) {};
		\node [style=mlpnode] (95) at (-3, 5.25) {};
		\node [style=mlpnode] (96) at (-3, 4.5) {};
		\node [style=mlpnode] (97) at (-3, 3.75) {};
		\node [style=mlpnode] (98) at (6, 2.75) {};
		\node [style=mlpnode] (99) at (6, 2) {};
		\node [style=mlpnode] (100) at (6, 1.25) {};
		\node [style=mlpnode] (101) at (6, 0.5) {};
		\node [style=none] (102) at (7, 2.75) {};
		\node [style=none] (103) at (7, 2) {};
		\node [style=none] (104) at (7, 1.25) {};
		\node [style=none] (105) at (7, 0.5) {};
		\node [style=mlpnode] (106) at (7, 2.75) {};
		\node [style=mlpnode] (107) at (7, 2) {};
		\node [style=mlpnode] (108) at (7, 1.25) {};
		\node [style=mlpnode] (109) at (7, 0.5) {};
	\end{pgfonlayer}
	\begin{pgfonlayer}{edgelayer}
		\draw [style=dashes] (64.center)
			 to [bend left=15] (65.center)
			 to (63.center)
			 to [bend left=15] (62.center)
			 to cycle;
		\draw [style=dashes] (68.center)
			 to (69.center)
			 to [bend left=15] (66.center)
			 to (67.center)
			 to [bend left=15] cycle;
		\draw (70) to (71.center);
		\draw (0) to (71.center);
		\draw [style=edge style 0] (71.center) to (4);
		\draw [style=edge style 0] (71.center) to (5);
		\draw [style=edge style 0] (71.center) to (6);
		\draw [style=edge style 0] (71.center) to (7);
		\draw (4) to (12);
		\draw (4) to (13);
		\draw (4) to (14);
		\draw (4) to (15);
		\draw (5) to (12);
		\draw (5) to (13);
		\draw (5) to (14);
		\draw (5) to (15);
		\draw (6) to (12);
		\draw (6) to (13);
		\draw (6) to (14);
		\draw (6) to (15);
		\draw (7) to (14);
		\draw (7) to (15);
		\draw (7) to (13);
		\draw (7) to (12);
		\draw (74) to (82);
		\draw (74) to (83);
		\draw (74) to (84);
		\draw (74) to (85);
		\draw (75) to (82);
		\draw (75) to (83);
		\draw (75) to (84);
		\draw (75) to (85);
		\draw (76) to (82);
		\draw (76) to (83);
		\draw (76) to (84);
		\draw (76) to (85);
		\draw (77) to (84);
		\draw (77) to (85);
		\draw (77) to (83);
		\draw (77) to (82);
		\draw (82) to (97);
		\draw (82) to (96);
		\draw (82) to (95);
		\draw (82) to (94);
		\draw (83) to (94);
		\draw (83) to (95);
		\draw (83) to (96);
		\draw (83) to (97);
		\draw (84) to (95);
		\draw (84) to (96);
		\draw (84) to (97);
		\draw (84) to (94);
		\draw (85) to (94);
		\draw (85) to (95);
		\draw (85) to (96);
		\draw (85) to (97);
		\draw [style=edge style 0] (94) to (32);
		\draw [style=edge style 0] (95) to (32);
		\draw [style=edge style 0] (96) to (32);
		\draw [style=edge style 0] (97) to (32);
		\draw (33) to (41);
		\draw (33) to (42);
		\draw (33) to (43);
		\draw (33) to (44);
		\draw (34) to (41);
		\draw (34) to (42);
		\draw (34) to (43);
		\draw (34) to (44);
		\draw (35) to (42);
		\draw (35) to (41);
		\draw (35) to (43);
		\draw (35) to (44);
		\draw (36) to (43);
		\draw (36) to (44);
		\draw (36) to (42);
		\draw (36) to (41);
		\draw (98) to (106);
		\draw (98) to (107);
		\draw (98) to (108);
		\draw (98) to (109);
		\draw (99) to (106);
		\draw (99) to (107);
		\draw (99) to (108);
		\draw (99) to (109);
		\draw (100) to (107);
		\draw (100) to (106);
		\draw (100) to (108);
		\draw (100) to (109);
		\draw (101) to (108);
		\draw (101) to (109);
		\draw (101) to (107);
		\draw (101) to (106);
		\draw (41) to (98);
		\draw (41) to (99);
		\draw (41) to (100);
		\draw (41) to (101);
		\draw (42) to (98);
		\draw (42) to (99);
		\draw (42) to (100);
		\draw (42) to (101);
		\draw (43) to (98);
		\draw (43) to (99);
		\draw (43) to (100);
		\draw (43) to (101);
		\draw (44) to (101);
		\draw (44) to (100);
		\draw (44) to (99);
		\draw (44) to (98);
		\draw [style=edge style 0] (32) to (33);
		\draw [style=edge style 0] (32) to (34);
		\draw [style=edge style 0] (32) to (35);
		\draw [style=edge style 0] (32) to (36);
		\draw [style=edge style 0] (1) to (33);
		\draw [style=edge style 0] (1) to (34);
		\draw [style=edge style 0] (1) to (35);
		\draw [style=edge style 0] (1) to (36);
		\draw [style=edge style 0] (2) to (33);
		\draw [style=edge style 0] (2) to (34);
		\draw [style=edge style 0] (2) to (35);
		\draw [style=edge style 0] (2) to (36);
		\draw [style=edge style 0] (3) to (36);
		\draw [style=edge style 0] (3) to (35);
		\draw [style=edge style 0] (3) to (34);
		\draw [style=edge style 0] (3) to (33);
		\draw [style=edge style 0] (106) to (61);
		\draw [style=edge style 0] (107) to (61);
		\draw [style=edge style 0] (108) to (61);
		\draw [style=edge style 0] (109) to (61);
	\end{pgfonlayer}
\end{tikzpicture}